\documentclass[onecolumn,longbibliography,superscriptaddress,
amsmath,amssymb, aps, pre]{revtex4-2}
\usepackage{dcolumn}
\usepackage{bm}
\usepackage{graphicx}
\usepackage{comment}
\usepackage[caption=false]{subfig}
\usepackage{color}
\usepackage{algorithm}
\usepackage{algpseudocode}
\usepackage{textcomp}

\newcommand{\bea}{\begin{eqnarray}}

\newcommand{\eea}{\end{eqnarray}}

\begin{document}

\begin{abstract}
Random walks on lattices with preferential relocation to previously visited sites provide a simple framework for modeling the displacements of animals and humans. When the lattice contains a few impurities or resource sites where the walker spends more time on average at each visit than on the other sites, the long range memory can suppress diffusion and induce by reinforcement a steady state localized around a resource. This phenomenon can be identified with a spatial learning process. 
Here we study theoretically and numerically how the decay of memory impacts learning in a model with one impurity. If memory decays as $1/\tau$ or slower, where $\tau$ is the time backward into the past, the localized solution is the same as with perfect, non-decaying memory and it is linearly stable. If forgetting is faster than $1/\tau$, for instance exponential, an unusual regime of intermittent localization is observed, where well localized periods of exponentially distributed duration are disrupted by possibly long intervals of diffusive motion. At the transition between the two regimes, for a kernel in $1/\tau$, the approach to the stable localized state is the fastest, opposite to the expected critical slowing down effect. Hence, forgetting can allow the walker to save a lot of memory without compromising learning and to achieve a faster learning process. These findings agree with biological evidence on the benefits of forgetting.
\end{abstract}

\preprint{}
\title{Intermittent localization and fast spatial learning by non-Markov random walks with decaying memory}
\author{Paulina R. Mart\'\i n-Cornejo}
\author{Denis Boyer}
\affiliation{Instituto de F\'\i sica, Universidad Nacional Aut\'onoma de M\'exico,
Ciudad de M\'exico 04510, M\'exico}

\maketitle

\section{Introduction}

Whereas the lattice random walk represents a paradigm for studying diffusion, some random walk models exhibit localization properties that are in or out-of-equilibrium \cite{golosov1984localization, davis1990reinforced,albeverio1998asymptotics,burda2009localization}.
A classical example is the so-called vertex reinforced random walk, consisting of a walker that hops to nearest-neighbor (n.n.) sites, such that the probability of transiting to a site depends linearly on the number of previous visits received by that site in the past \cite{pemantle1988phase,pemantle1992vertex,pemantle1999vertex}. As the walker needs to keep track of all the visits to all the sites, the process has unbounded memory and is highly non-Markov unlike the ordinary random walk. A consequence of local attraction toward sites that are visited more frequently can be the suppression of diffusion and the emergence of localization. The reinforced random walk in one dimension ($1d$), for instance, exhibits a highly non-trivial property: it gets stuck asymptotically with probability 1 on five sites of the lattice, that are visited infinitely often \cite{tarres2004vertex}.

Other models based on the principle of preferential returns to previously visited locations have been developed over the  past two decades and quantitatively compared to real movement data of single animals 
\cite{gautestad2005intrinsic,boyer2014random,ranc2021experimental,vilk2022phase} or humans \cite{song2010modelling}. Empirical observations actually indicate that individuals across many species use their memory and tend to return preferentially to familiar places in their environment \cite{fagan2013spatial}, i.e., that the probability of revisiting a location is roughly proportional to the total amount of time spent at this location previously \cite{song2010modelling,boyer2014random}. It is also well established that most animals have home ranges, i.e., are localized on a rather limited area instead of diffusing randomly and unbounded. The mechanisms of home range formation are still not fully understood \cite{borger2008there}, but memory is assumed to play a central role \cite{ranc2022memory}. However, most ecological movement models based on preferential revisits, although capturing realistic features of field studies, do not commonly exhibit localization but rather very slow diffusion. 

An analytically solvable example is the preferential relocation model or the 'monkey walk' \cite{boyer2014random,mailler2019random,boci2023large,boci2024central}, which mixes random exploration with self-attraction. At each discrete time-step, a walker on an infinite $1d$ lattice takes a usual random n.n. move with probability $1-q$ or resets to a previous site (say $j$) with the complementary probability $q$. In the latter eventuality, the site $j$ is chosen among all visited sites with a probability proportional to the total amount of time spent there since the beginning of the process. One important difference from the classic reinforced random walk lies in the fact that the chosen site for a revisit does not need to be a nearest neighbor of the walker's current position. For any non-zero $q$, it is found that the occupation probability is always time-dependent and obeys a Gaussian scaling-law at large times, but with a variance growing extremely slowly, as $\ln (t)$. In other words, the process does not converge towards a stationary state (which would indicate localization), but evolves much more slowly than in normal diffusion, where the mean square displacement increases as $t$. 

A simple modification of the above model does exhibit localization, though. Assume that one particular site, placed at the origin, is more attractive than the other ones: each time the walker occupies the origin, it stays there for one more time-step with probability $\gamma$ and does the monkey walk with probability $1-\gamma$. On the other sites, the walker only performs monkey walk moves. This setup mimics an animal that occasionally feeds on a particular resource site before continuing its motion in a scarce environment. In the limit $t\to\infty$, analytical arguments show that, in $1d$, a localized solution independent of time exists for any non-zero values of the probabilities $q$ and $\gamma$ \cite{FBGM2017}. The steady state occupation probability is centered around the resource site and is no longer Gaussian but with exponential tails, which define a finite localization length. If the underlying Markov process between resetting events is not recurrent, e.g., the walker performs a $1d$ L\'evy flight with an index smaller than unity, or a n.n. random walk in $3d$, the process localizes only if $q>q_c$ with $q_c>0$. Overall, the description of this phase transition is very similar phenomenologically to the self-consistent theory of Anderson localization \cite{FBGM2017,BFGM2019}. Extensions of this model to several resource sites exhibit similar features, including the enhancement of localization on the stronger impurities \cite{sepehrinia2025localization}.

There exists a fundamental difference between this type of localization and the ones observed in the classic reinforced random walk, or in the Markov random walk with resetting to the starting position only \cite{evans2011diffusion,evans2011diffusionb,evans2020stochastic,gupta2022stochastic,pal2015diffusion,eule2016non,pal2019time,bodrova2020resetting}: the steady state is independent of the initial position and represents an {\em attractor} of the dynamics. This attractor is largely determined by the medium itself (the position and attractiveness $\gamma$ of the resource site) and results from the experience of the walker accumulated over time. At late times, the walker settles down around the impurity, in a way actually very similar to a memory-less random walker with resetting to the impurity only \cite{FBGM2017,BFGM2019}, but without knowing before hand the existence of this convenient resetting position on the lattice. Hence, convergence toward localization can be identified as a {\em learning process}, in the sense understood in the behavioral sciences, i.e., a change in behavior resulting from experience \cite{seifert2019learning}. 

The purpose of the present study is to analyze the effects of a decaying memory on the existence of localized solutions and on their stability, in the learning monkey walk problem described above. Random walk models with preferential revisits usually assume that memory does not decay with time, or all visits are counted equally. Assuming that memory is not perfect, e.g., that more recent visits are remembered better than those of the remote past, is  more realistic biologically. Exponential or power-law memory kernels have been identified, for instance, by fitting animal ranging data to models with memory decay \cite{merkle2014memory,falcon2021hierarchical,ranc2021experimental}. In preferential relocation processes, the main effect of memory decay is a possible modification, on homogeneous lattices, of the logarithmic diffusion law to a faster one, in $t^{\mu}$ with $0<\mu\le 1$ \cite{boyer2014solvable,mailler2019random,boyer2017long}, or changes in their relaxation dynamics under confinement \cite{boyer2025diffusion}. 

In the neurosciences, forgetting, apart from being unavoidable, does not always represent a dysfunction of the brain but has some virtues:  it contributes to suppress interferences between different memories and makes room for retaining new information \cite{markovitch1988role,wixted2005theory,hardt2013decay}. From the modeling side, mean-field approaches show that exponential forgetting can actually help foraging animals to cope with changing environments \cite{mcnamara1985optimal}. In numerical simulations of a swarm of interacting monkey walks in a static but heterogeneous environment containing many impurities, the decay of memory was seen to improve the aggregation of individuals around the best resource sites, a phenomenon identified as a successful collective learning \cite{falcon2023levy}. Very few analytical results are available on the relation between forgetting and spatial learning. To start with, is perfect memory a necessary condition for localization in preferential random walk models?

We now present the structure of the article and summarize the main results. In Section \ref{sec:model} we introduce the memory random walk model with one impurity on the infinite lattice, where forgetting is incorporated via a general kernel. In Section \ref{sec:perfectmemory}, we recall known analytical results on the steady state solution of the problem with perfect memory. Section \ref{sec:numres} reports simulation results. When memory has an asymptotic decay slower than $1/\tau$ (where $\tau$ is the time backward into the past), i.e., of the form $1/\tau^{\beta}$ with $0\le\beta\le1$, we find that the system tends toward the same localized state at late times than with perfect memory or $\beta=0$ (Section \ref{sec:slowmem}). This fact is consistent with the prediction obtained from a decoupling approximation. However, with kernels decaying faster than $1/\tau$ (Sections \ref{sec:expmem} and \ref{sec:1beta2}), an unusual dynamical behavior is observed: localization becomes intermittent in time, or spatial learning is disrupted, i.e., well-localized periods of finite durations are separated by possibly very long intervals of diffusive motion. Somehow surprisingly, during the localized periods, the occupation probability of the lattice sites is very close to the solution with perfect memory. In Section \ref{sec:stability}, we perform a linear stability analysis of the fully localized solution using the decoupling approximation, which turns out to be consistent only for memory kernels decaying more slowly than $1/\tau$.  In these cases, we find that the localized solutions are stable, in agreement with the numerics of Section \ref{sec:slowmem}. We then study the relaxation at late times toward these stable states (Section \ref{sec:rel}), showing that it is non-exponential and of the form $t^{\nu}/\ln t$, with $\nu\in(-1,0)$ a non-trivial exponent solution of Eq. (\ref{nu}). In the boundary case $\beta=1$ corresponding to the $1/\tau$ memory kernel at the edge of intermittency, the localized state is still stable and the relaxation towards the latter is actually the {\em fastest} among all the values of $\beta\in[0,1]$: the amplitudes of the perturbations tend to zero no more slowly than $(\ln t)^{\alpha}/t$, with $\alpha$ an exponent [Eqs. (\ref{ansatz2})-(\ref{alpha})]. This speed-up, confirmed by numerical simulations, is opposite to the usual critical slowing down for a state that is marginally stable. Hence, the $1/\tau$ forgetting law is in some sense optimal, as $(i)$ it achieves the same localization as perfect memory by being the most economical on memory, and $(ii)$ it localizes the fastest. In Section \ref{sec:scenarios} we discuss different possible scenarios for intermittent localization (or imperfect learning) and conclude in Section \ref{sec:concl}.

\section{The model}\label{sec:model}

We revisit the discrete time random walk model with preferential relocation to previously visited sites, in the presence of one impurity \cite{FBGM2017}. The original model considers a $1d$ lattice and a random walker with perfect memory that performs symmetric n.n. random steps with probability $1-q$, and resets to a previously visited site with probability $q$. In the latter case, a site is chosen, among the visited sites, with a probability proportional to the total amount of time spent on that site since $t=0$. Additionally, an impurity site is located at the origin $n=0$, representing a food resource site. When the walker, with position $X_t$ at time $t$, is located at the origin (or $X_t=0$), it stays on that site for the next time-step with probability $\gamma$ and moves according to the two movement rules above with the complementary probability $1-\gamma$. Hence, the parameter $0\le\gamma\le 1$ represents the attractiveness of the origin, which is occupied for a longer time at each visit compared to the other sites of the lattice. This site is likely to be more strongly reinforced by the resetting dynamics.

As mentioned above, when memory is used during the time-step $t\to t+1$, a previous site is chosen to be revisited with a probability proportional to the accumulated amount of time spent there. This linear rule is equivalent to choosing a time $t'$ in the past, $t'=0,1,...,t$, with uniform probability and to relocate to the position occupied at time $t'$, i.e., to set 
\begin{equation}\label{resetmove}
X_{t+1}=X_{t'}\,. 
\end{equation}
A site that has been often occupied will have more possibilities (times $t'$) of being chosen. The fact that $t'$ is distributed uniformly also means that the walker remembers equally well all its previous positions, or memory does not decay. 

To incorporate memory decay in this model, let us consider a more general probability distribution for $t'$, where more recent time-steps are remembered better than the initial ones. To this end, at time $t$, the time $t'$ in the past that appears in Eq. (\ref{resetmove}) is chosen with probability $\pi_{t,t'}$ and the memory kernel is of the form $\pi_{t,t'}\propto F(t-t')$ with $F(\tau)$ a decreasing function of its argument. Since $\sum_{t'=0}^t \pi_{t,t'}=1$ by normalization, one has
\begin{equation}\label{pi}
\pi_{t,t'}=\frac{F(t-t')}{\sum_{\tau=0}^t F(\tau)}\,.
\end{equation}
Setting $F(0)=1$ by convention, the quantity in the denominator
\begin{equation}\label{C}
C(t)=\sum_{\tau=0}^t F(\tau)\, ,
\end{equation}
represents the effective number of time-steps that are remembered by the walker at time $t$.
The linear preferential rule studied in \cite{FBGM2017} corresponds to the uniform distribution $F(\tau)=1$, or perfect memory case, which gives $C(t)=t+1$ (all visits are remembered) and $\pi_{t,t'}=1/(t+1)$.

Let us denote as $P_n(t)$ the walker's occupation probability of the site $n$ at time $t$, $P_n(t)={\rm Prob}[X_t=n]$, given an unspecified initial condition. This quantity satisfies the following master equation,
\begin{eqnarray}
\label{ME}
P_n(t+1)&=&\frac{1-q}{2}\left[(1-\gamma\delta_{n+1,0})P_{n+1}(t)+ (1-\gamma\delta_{n-1,0})P_{n-1}(t)\right]
+\gamma\delta_{n,0}P_n(t)\\
&+&q(1-\gamma)\sum_{t'=0}^{t}\pi_{t,t'}{\rm Prob}[X_{t'}=n\ {\rm and}\ X_t=0]\nonumber\\
&+&q\sum_{t'=0}^{t}\pi_{t,t'}{\rm Prob}[X_{t'}=n\ {\rm and}\ X_t\neq 0]\,.\nonumber
\end{eqnarray}
The first line of Eq. (\ref{ME}) represents the Markov random walk steps and trapping by the impurity, where $\delta_{i,j}$ is the Kronecker function. The second line accounts for the probability of revisiting the site $n$ via memory at time $t+1$ (because it was visited at a previous time $t'$) by jumping from the origin which is occupied at time $t$. The third line counts the relocation jumps to $n$ from another site than the origin. The key point in this problem is that the resetting probability actually depends on the position: it is $q(1-\gamma)$ at the origin (since the walker does not move with probability $\gamma$) and $q$ at the other sites. Such heterogeneity is crucial for the existence of localized solutions, but it also makes the problem very difficult to solve, as Eq. (\ref{ME}) cannot be reduced to a closed form involving the marginal distribution $P_n(t)$ only. Clearly, the master equation explicitly depends on higher-order joint probabilities such as ${\rm Prob}[X_{t'}=n\ {\rm and}\ X_t=0]$, therefore a whole hierarchy should be written for the multiple-time distribution functions \cite{BFGM2019}.

\section{Previous results for the perfect memory case}
\label{sec:perfectmemory}

We summarize the results obtained from the uniform memory kernel $\pi_{t,t'}=1/(t+1)$, corresponding to no memory decay. 

\subsection{Homogeneous lattice ($\gamma=0$)}

For completeness, let us first recall the main properties of the monkey walk on a homogeneous lattice free of impurity, corresponding to $\gamma=0$ \cite{boyer2014random,boyer2017long,mailler2019random}. This problem is more tractable since Eq. (\ref{ME}) closes in that case and can be recast as,
\begin{equation}
P_n(t+1)=\frac{1-q}{2}\left[P_{n+1}(t)+P_{n-1}(t)\right]+\frac{q}{t+1}\sum_{t'=0}^t P(n,t')\, .
\end{equation}
This equation is linear and the mean-square displacement of a particle starting from the origin, $M_2(t)\equiv\sum_{n=-\infty}^{\infty}n^2P(n,t)$, can be obtained exactly \cite{boyer2014random},
\begin{equation}
M_2(t)=\frac{1-q}{q}\sum_{k=1}^t \frac{1-(1-q)^k}{k}\simeq \frac{1-q}{q}\left[\ln(qt)\,+\,\gamma+{\rm h.o.t.}\right],
\end{equation}
where $\gamma$ is the Euler constant and h.o.t. represents higher order terms, decaying as $1/t$ at large $t$. Hence, diffusion is logarithmically slow in time but unbounded. At large $t$, the distribution $P_n(t)$ obeys a scaling form,
\begin{equation}\label{scaling}
P_n(t)\simeq \frac{1}{\phi(t)} G\left(\frac{n}{\phi(t)}\right)\, ,
\end{equation}
with $\phi(t)=\sqrt{M_2(t)}$ and $G(z)=\frac{1}{\sqrt{2\pi}}e^{-\frac{z^2}{2}}$ the centered Gaussian distribution of unit variance. The occupation probability in Eq. (\ref{scaling}) is therefore always time-dependent, tends to $0$ uniformly at late times and no localized solutions exist.

\subsection{Localization with $\gamma\neq0$}

\begin{figure}[t]
    	\begin{center}
            \subfloat[]{%
            \includegraphics[width = 0.47\textwidth]{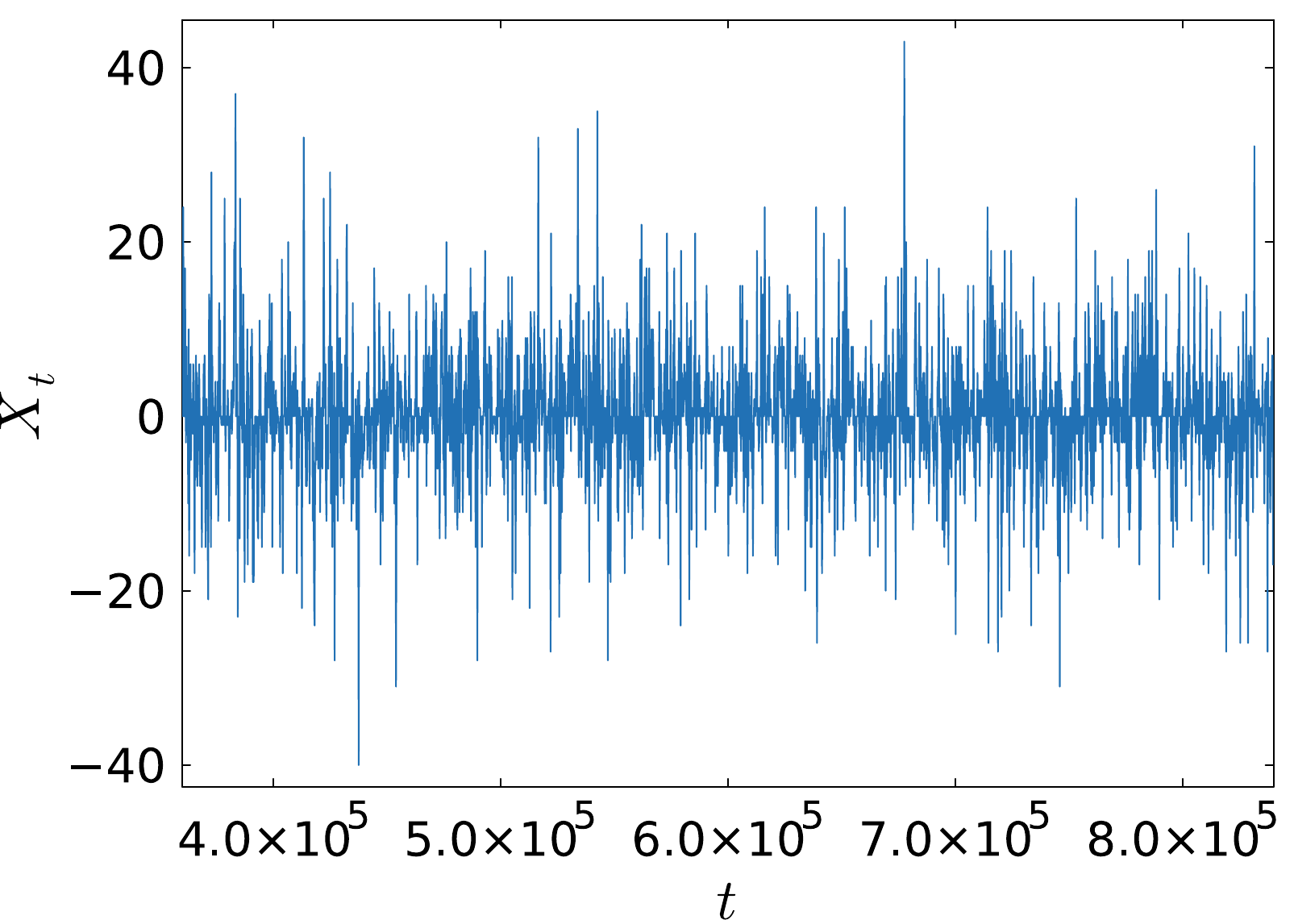}
            } \hfil
            \subfloat[]{%
            \includegraphics[width = 0.47\textwidth]{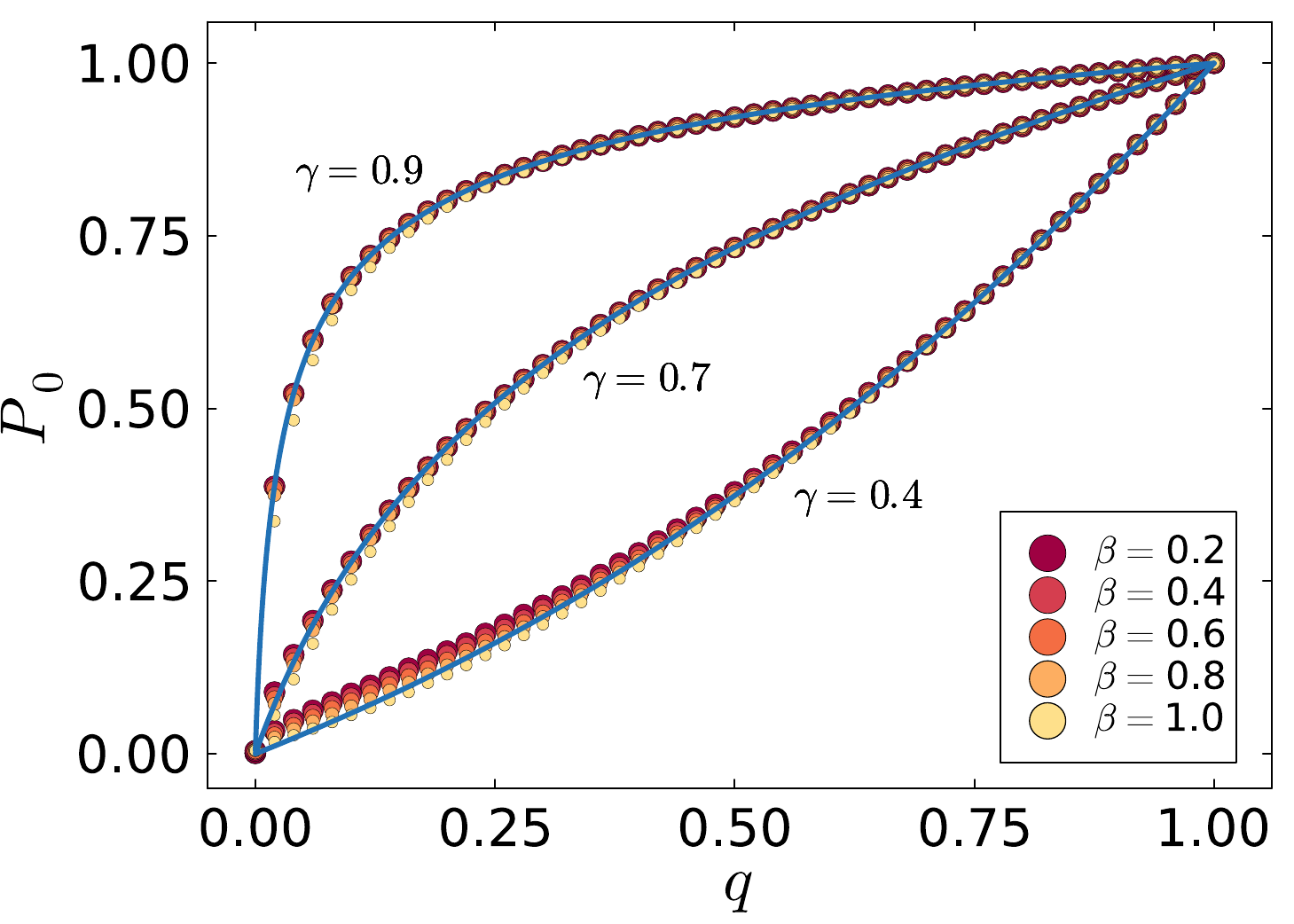}
            }
	\end{center}
    \caption{(a) A localized trajectory $X_{t}$ vs. {t} with power-law memory decay, for $\beta = 1$,  obtained from a numerical simulation with $q = 0.1$ and $\gamma=0.7$. (b) Occupation probability of the origin, $P_{0}$, as a function of $q$ for various values of $\gamma$. Solid (blue) lines are given by Eq. (\ref{P0}) and symbols by Monte Carlo simulations ($t = 10^{6}$) corresponding to different memory decay exponents $\beta \in [0, 1]$.}
    \label{fig:powerlaw}
\end{figure}

In this case, let us assume that $P_n(t)$ in Eq. (\ref{ME}) tends to a stationary distribution $P_n^{(st)}> 0$ at large times for all $n$, peaked around the origin \cite{FBGM2017}. This solution, if it exists, corresponds to a localized phase. Some progress can be made in the resolution of Eq. (\ref{ME}) with $\gamma\ne0$ by noticing that the times $t'$ and $t$ are in general very far apart asymptotically, justifying the use of a de-correlation approximation,
\begin{eqnarray}
&& {\rm Prob}[X_{t'}=n\ {\rm and}\ X_t=0]\approx {\rm Prob}[X_{t'}=n]\times{\rm Prob}[X_t=0]= P_n(t')P_0(t)\label{decoupling}\\
&& {\rm Prob}[X_{t'}=n\ {\rm and}\ X_t\neq 0]\approx {\rm Prob}[X_{t'}=n]\times{\rm Prob}[X_t\neq 0]= P_n(t')[1-P_0(t)]\, .\label{decouplingb}
\end{eqnarray}
Next, we assume that the probabilities can be replaced by their stationary values
\begin{eqnarray}\label{steady}
&&P_n(t')\approx P_n^{(st)}\nonumber\\
&&P_0(t)\approx P_0^{(st)}\\
&&P_n(t)\approx P_n^{(st)}.\nonumber
\end{eqnarray}
Inserting these expressions into Eq. (\ref{ME}) gives the nonlinear equation,
\begin{equation}\label{MEsteady}
P_n^{(st)}\approx\frac{1-q}{2}[P_{n+1}^{(st)}+ P_{n-1}^{(st)}]+\gamma P_0^{(st)}\left[\delta_{n,0}-\frac{1-q}{2}(\delta_{n,1} +\delta_{n,-1})\right]+qP_n^{(st)}[1-\gamma P_0^{(st)}]\,.
\end{equation}
This equation can be exactly solved for the Fourier transform of $P_{n}^{(st)}$, where the occupation probability of the origin $P_0^{(st)}$ is considered as a constant that is finally obtained self-consistently. The analytical expression for the occupation probability at the origin was obtained in \cite{FBGM2017} and reads
\begin{equation}\label{P0}
 P_0^{(st)} = \frac{-(1-q)(1-\gamma)^2-q\gamma^2}{q\gamma(1-2\gamma)}
     + \frac{\sqrt{[(1-q)(1-\gamma)^2+q\gamma^2]^2+(q\gamma)^2(1-2\gamma)}}{q\gamma(1-2\gamma)}\, ,
\end{equation}
for $\gamma\ne1/2$, while $P_0^{(st)}=q$ for $\gamma=1/2$.
The distribution decays exponentially with the distance $|n|$ to the origin and is given by 
\begin{equation}\label{Pn}
 P_n^{(st)}=\gamma P_0^{(st)}\delta_{n,0}+(1-\gamma)P_0^{(st)}a^{-|n|}\,,
\end{equation}
with 
\begin{eqnarray}
a&=&1+u+\sqrt{u\left(2+u\right)}>1\,,\label{a}\\
u&=&\frac{\gamma q P_0^{(st)}}{1-q}\,.\label{u}
\end{eqnarray}
A quick analysis of Eq. (\ref{P0}) shows that $P_0^{(st)}>0$ for all $q>0$ and $\gamma>0$. Consequently, the localization length $\xi=(\ln a)^{-1}$ is always finite and the walker localizes for any non-zero parameter values $q$ and $\gamma$, a fact that was checked in numerical simulations of the model \cite{FBGM2017}.  In addition, the agreement of Eqs. (\ref{P0})-(\ref{Pn}) with numerical simulations was very good (as also illustrated by Fig. \ref{fig:powerlaw}b below), suggesting that the de-correlation approximation above might be exact.
With the solution (\ref{P0})-(\ref{Pn}) at hand, it is also possible to calculate the participation ratio, a quantity of interest to characterize localization in classical \cite{compte1998localization}  
and quantum systems \cite{hikami1986localization,murphy2011generalized, calixto2015inverse}. Here, it is defined as the probability that two independent walkers occupy the same site,  
\begin{equation}
PR=\sum_{n=-\infty}^{\infty}P_n^2\, ,
\end{equation}
and it is given by the expression
\begin{equation}\label{PR}
PR=\left[1+2\frac{(1-\gamma)^2}{a^2-1}\right]P_0^2\,.
\end{equation}
Hence $PR$ is finite and does not decay to zero with the system size as in ordinary diffusion.

Localization in this context can be interpreted as a phenomenon of spatial learning, where the walker \lq\lq learns" to exploit the resource patch at the origin instead of diffusing far away from it. As shown below in Fig. \ref{fig:powerlaw}b, localization is stronger ($P_0^{(st)}$ larger) if the resource patch is more attractive ($\gamma$ large) or memory is more frequently used ($q$ large). Interestingly, on $3d$ lattices, the same de-correlation approximation predicts that a localization/delocalization phase transition takes place at $q=q_c$ with $q_c$ a non-zero critical value, in agreement with simulations \cite{BFGM2019}. In addition, the localization length $\xi$ diverges near $q_c$ with the same critical exponents as in the self-consistent theory of Anderson localization \cite{FBGM2017,BFGM2019}. If the n.n. random walk is replaced by a L\'evy flight with sufficiently small index, a localization/delocalization transition takes place on the $1d$ lattice as well \cite{FBGM2017}. In the $1d$ problem studied here, $q_c=0$ and there is no de-localized phase.

\section{numerical results with memory decay}\label{sec:numres}

We now address the case of a general memory kernel $\pi_{t,t'}$ and consider a resource site of attractiveness $\gamma>0$. If we apply the same approximations (\ref{decoupling})-(\ref{steady}) to Eq. (\ref{ME}), we obtain Eq. (\ref{MEsteady}) again. This is simply due to the fact that the normalization condition $\sum_{t'=0}^t\pi_{t,t'}=1$ is valid for all kernels. Hence, under the same assumptions, the results (\ref{P0})-(\ref{PR}) hold for {\em any kernel}, not just the perfect memory case. This, of course, cannot be correct if memory is very short-range, with $\pi_{t,t'}$ very peaked near the present time $t$, i.e., $F(\tau)$ tends rapidly to $0$ in Eq. (\ref{pi}). In that case, we expect the walker to behave roughly as a standard random walk with a modified diffusion coefficient, as shown in \cite{boyer2014solvable} in the absence of impurity. No stationary localized solutions should be expected in principle. 

To build intuition on the behavior of the system in the different cases, we have run Monte Carlo simulations generating stochastic trajectories with the rules of the model. Appendix \ref{sec:rndgen} gives details on the generation of the random times $t'$. We have considered the following choices for the function $F(\tau=t-t')$.

\subsection{Slowly decaying memory}\label{sec:slowmem}

In the first class of examples, the memory slowly decays as a power-law, and not faster than $1/\tau$. This type of law was observed in experiments on human subjects (the Ebbinghaus’ forgetting curve \cite{murre2015replication}) or inferred in the wild from the trajectories of large herbivores \cite{merkle2014memory}. We denote,
\begin{equation}\label{pipowerlaw}
F(\tau)=\frac{1}{(1+\tau)^{\beta}}\quad {\rm with}\quad 0\le \beta\le 1,
\end{equation}
where $\tau=0,1,...$ In this range of exponent values, the mean forgetting time, defined as \cite{boyer2014solvable}
\begin{equation}\label{forgettime}
\langle \tau\rangle= \lim_{t\to\infty}\frac{\sum_{t'=0}^t (t-t')F(t-t')}
{\sum_{t'=0}^t F(t-t')}\, ,
\end{equation}
is infinite. In addition, the effective number of visits that are remembered by the walker up to time $t$ [Eq. (\ref{C})] keeps increasing unbounded, albeit more slowly than $t$. Hence, memory is quite long-range for these kernels and the perfect memory (non-decaying) is recovered with $\beta=0$. 

We display in Fig. \ref{fig:powerlaw}a a typical trajectory $X_t$ vs. $t$ with $\beta=1$, showing a clear stationary localized behavior at late times, where the particle is never very far away from the origin. Figure \ref{fig:powerlaw}b shows the occupation probability $P_0$ computed at $t=10^6$ as a function of $q$ for various values of $\gamma$. In each case, the curves corresponding to different values of $\beta$ in the interval $[0,1]$ practically fall on top of each other and are in very good agreement with the theoretical prediction (\ref{P0}) of the de-correlation approximation. Hence, the applicability range of this approximation is not limited to the perfect memory studied in \cite{FBGM2017} and extends to kernels with sufficiently slow forgetting as well.

\begin{figure}[t]
    	\begin{center}
            \subfloat[]{%
            \includegraphics[width = 0.45\textwidth]{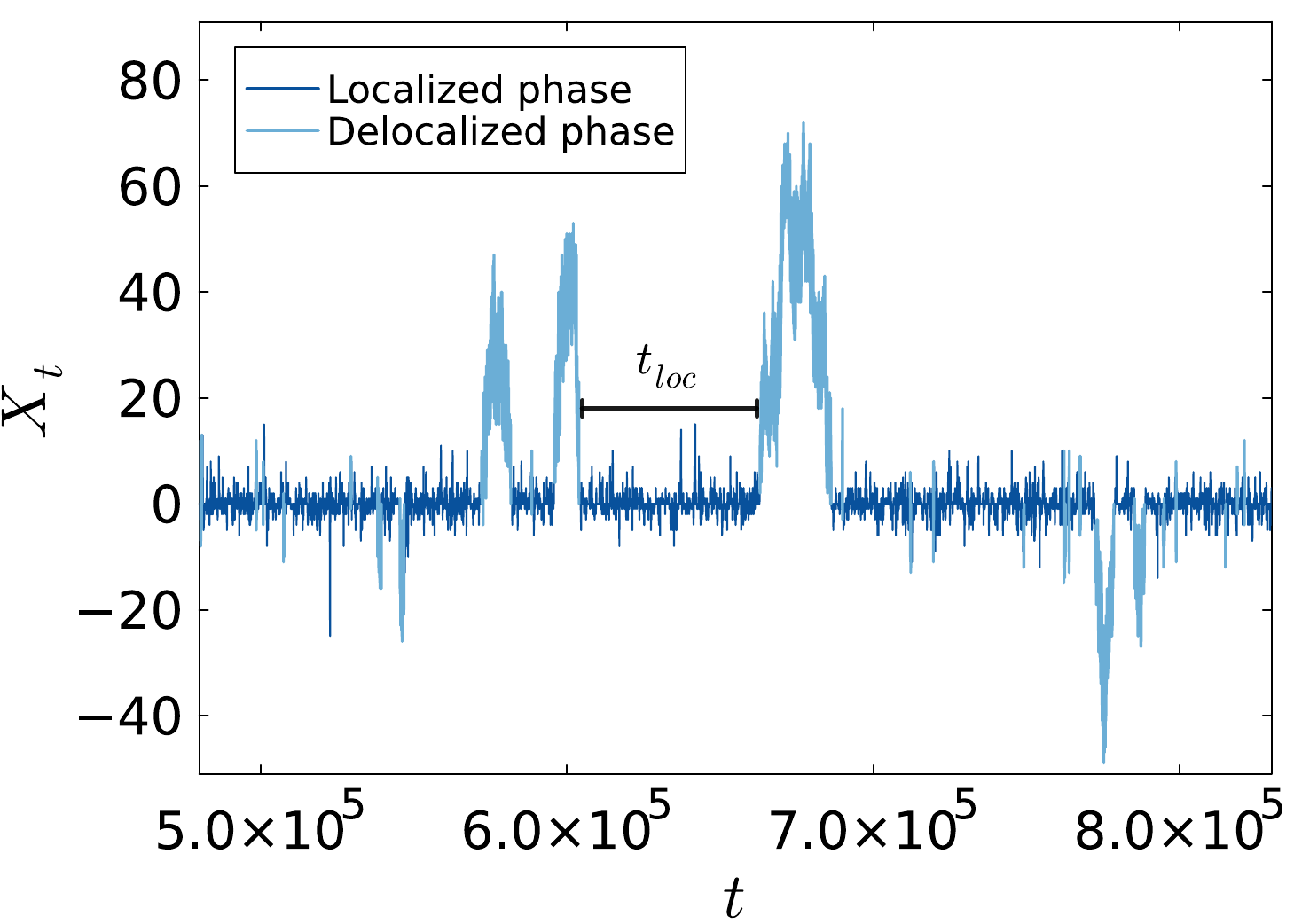}
            } \hfil
            \subfloat[]{%
            \includegraphics[width = 0.45\textwidth]{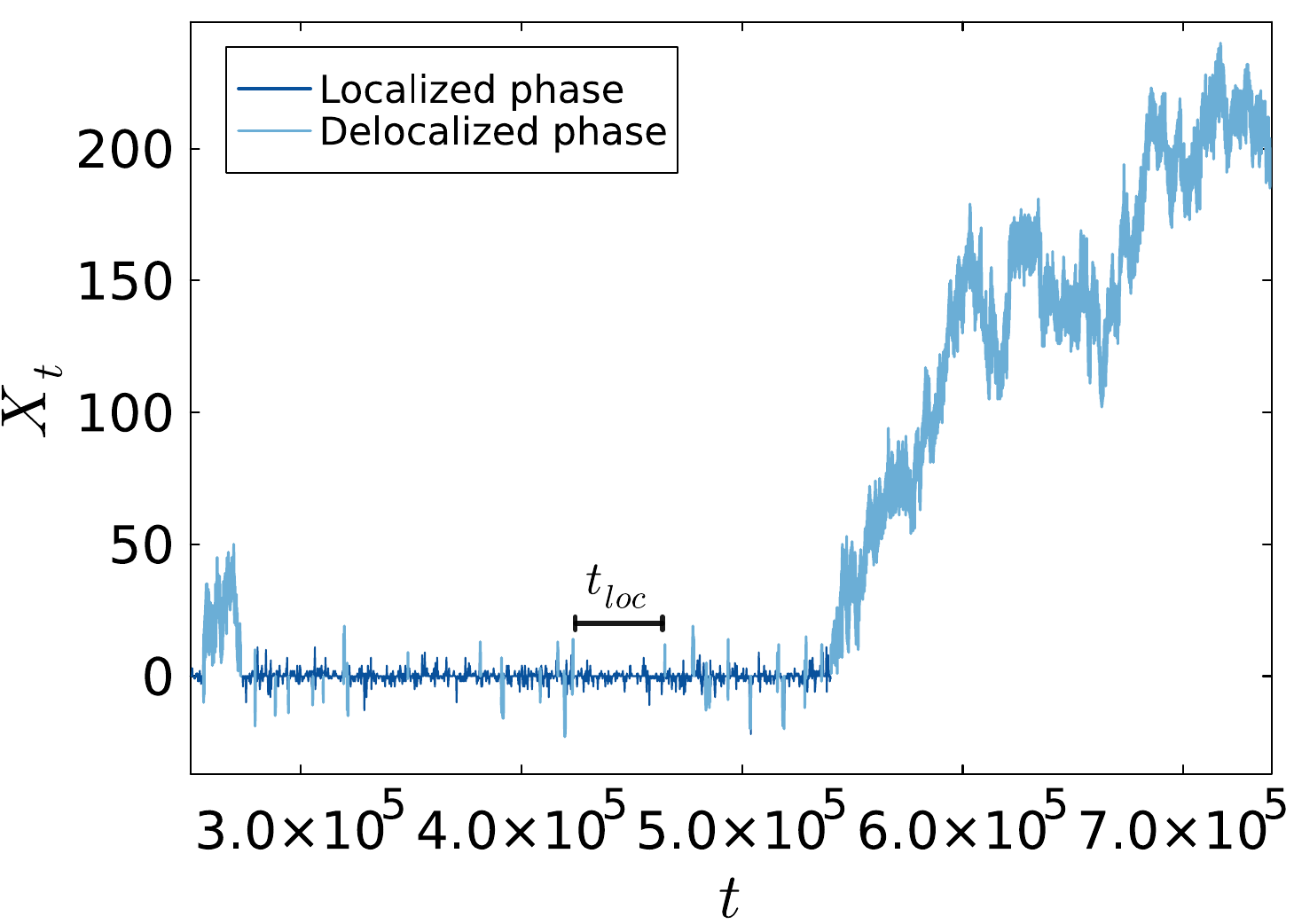}
            } \vfil
            \subfloat[]{%
            \includegraphics[width = 0.45\textwidth]{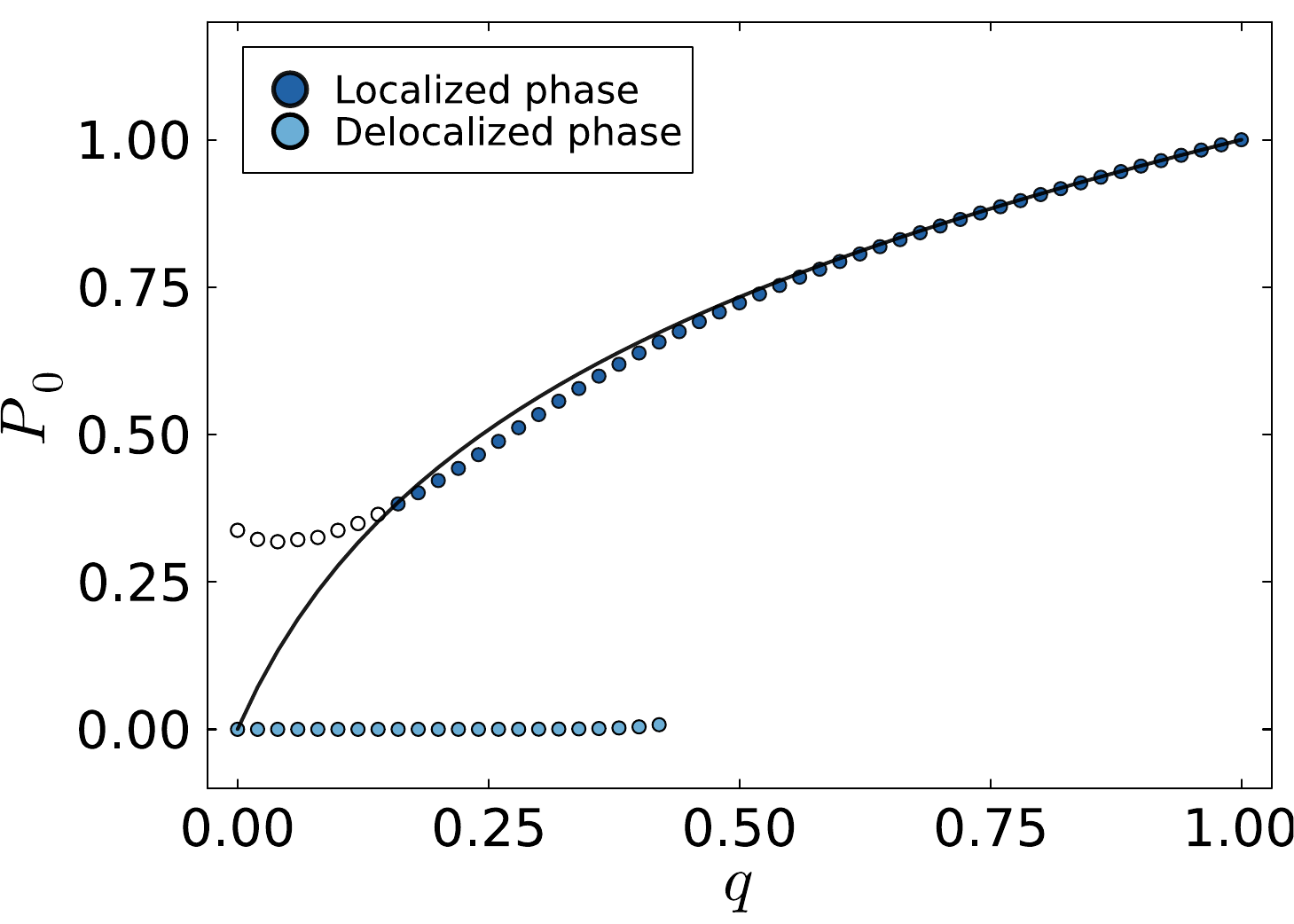}
            } \hfil
            \subfloat[]{%
            \includegraphics[width = 0.45\textwidth]{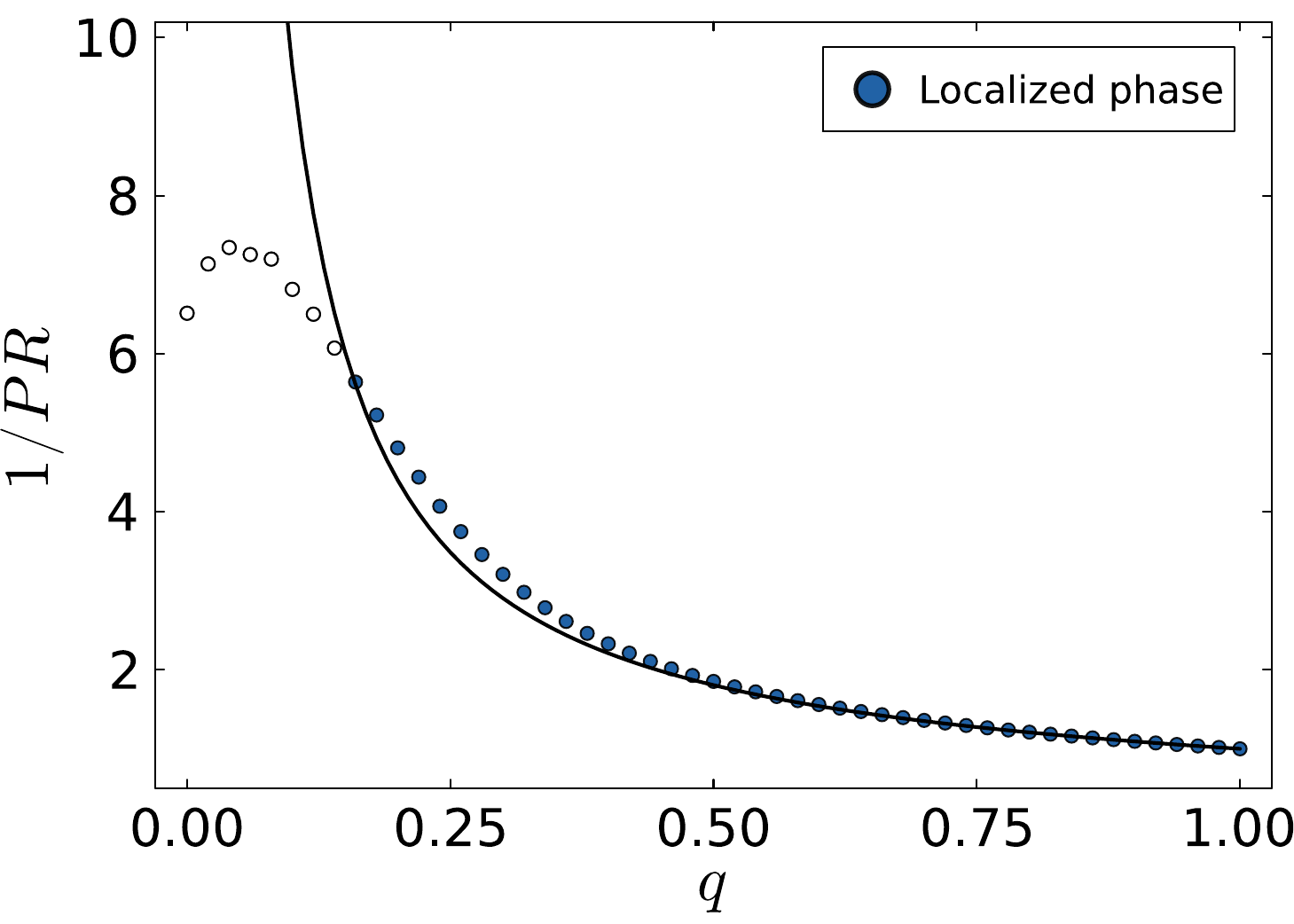}
            } 
	\end{center}
    \caption{(a)-(b) Two trajectories with exponential memory decay. The parameters are $\Delta = 30$, $q=0.3$ and $\gamma=0.7$. (c) Occupation probability of the origin in each phase and (d) inverse participation ratio of the localized phase. The solid lines are given by Eqs. (\ref{P0}) and (\ref{PR}), respectively.} 
    \label{fig:exponP0}
\end{figure}

\subsection{Exponentially decaying memory}\label{sec:expmem}

The next example, on the opposite, considers walkers with short range memory. A simple choice is the exponential decay \cite{mcnamara1985optimal}
\begin{equation}\label{Fexp}
F(\tau)=\exp\left(-\frac{\tau}{\Delta}\right)\,,
\end{equation}
with $\Delta$ representing a finite mean forgetting time $\langle \tau\rangle$. 

Two typical trajectories are displayed in Figs. \ref{fig:exponP0}a-b and a new type of behavior can be observed: the walker is not completely localized nor diffusive but stochastically alternates between these two types of motion. Fig. \ref{fig:exponP0}b, for instance, shows a very long diffusive phase at late times. We have identified the two phases using a simple algorithm described in Appendix \ref{sec:phasesep} and have analyzed them separately. Notably, the occupation probability $P_0$ at the origin computed from averaging only over the localized parts of many trajectories, shown in Fig. \ref{fig:exponP0}c, is practically undistinguishable from $P_0^{(st)}$ in Eq. (\ref{P0}). In comparison, the occupation probability $P_0$ in the de-localized phase is very low. (Note that if $q$ is too small, the identification of the two phases becomes difficult and cannot be done in a reliable way; these cases are labeled with open circles.)
Likewise, the inverse participation ratio obtained numerically from the distribution of the positions visited during the localized periods only is quite close to the theoretical prediction (\ref{PR}), provided that $q$ is not too small for the same reason as above (Fig. \ref{fig:exponP0}d).

These findings reveal that localization can be \lq\lq intermittent", i.e., effective during finite periods of time, albeit with practically the same properties as the fully localized solution. These features suggest a much wider applicability than expected of the analytical solution of Section \ref{sec:perfectmemory}. Although the de-correlation assumption might work poorly at large times for a kernel such as (\ref{Fexp}), it seems to be quite relevant at intermediate scales.

\begin{figure}[t]
    	\begin{center}
            \subfloat[]{%
            \includegraphics[width = 0.45\textwidth]{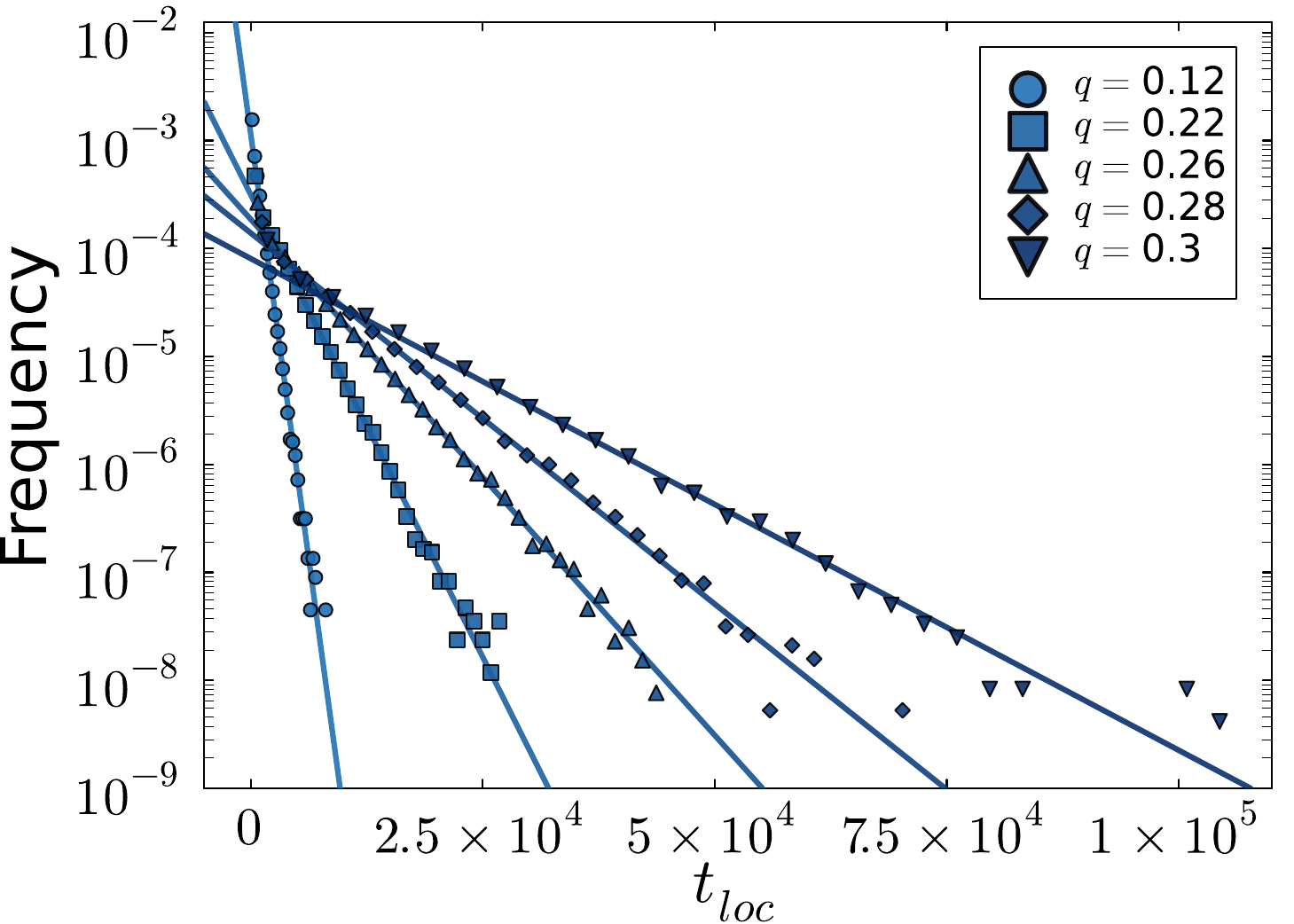}
            } \hfil
            \subfloat[]{%
            \includegraphics[width = 0.45\textwidth]{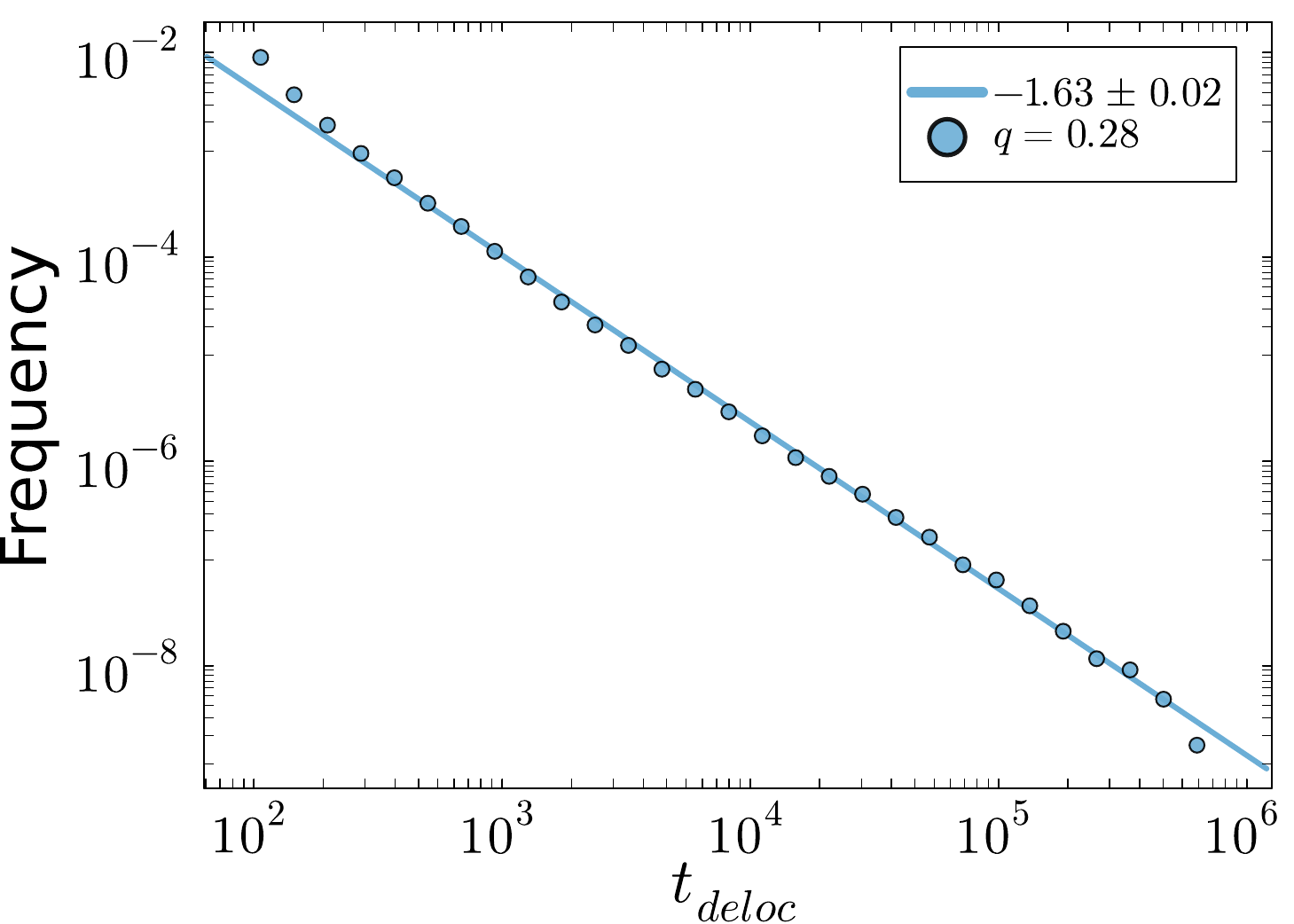}
            }
	\end{center}
    \caption{(a) Distributions of the lifetime $t_{loc}$ of localized periods. Symbols are given by Monte Carlo simulations ($t = 10^{6}$) corresponding to different values of $q$, for $\gamma = 0.7$ and $\Delta = 30$. The solid (dark blue) lines are fits with Eq. (\ref{plife}). (b) Distributions of the lifetime $t_{deloc}$ of delocalized periods. Symbols are given by Monte Carlo simulations ($t = 10^{6}$) corresponding to $q = 0.28$, $\gamma = 0.7$ and $\Delta = 30$ and the solid (light blue) line by fitting Eq. (\ref{rholife}).}
    \label{fig:lifetime}
\end{figure}

Intermittent localization can be further characterized by the lifetime of the localized periods, denoted as $t_{loc}$ and which is illustrated in Figs. \ref{fig:exponP0}a-b. This time is a random variable whose distribution $p(t_{loc})$ is computed numerically in Fig. \ref{fig:lifetime}a for several values of $q$. All the distributions are very well fitted by exponentials, 
\begin{equation}\label{plife}
p(t_{loc})\propto \exp\left(-\frac{t_{loc}}{\tau_0}\right)\,,
\end{equation}
practically over the whole range of lifetimes $t_{loc}$. Despite the non-Markov nature of the dynamics, $p(t_{loc})$ is remarkably simple and similar to the statistics of a Markov process, such as the escape from a potential well.
In contrast, the durations $t_{deloc}$ of the delocalized periods follow a very different distribution. In Fig. \ref{fig:lifetime}b, it is well fitted at large times by an inverse power-law 
\begin{equation}\label{rholife}
\rho(t_{deloc})\sim t_{deloc}^{-\alpha}\, ,
\end{equation}
with an exponent $\alpha$ close to $3/2$. This scaling is expected from the recurrence properties of simple $1d$ Polya random walks, where the time intervals $t$ between two consecutive visits of the origin have probability $t^{-3/2}$ at large $t$ \cite{krapivsky2010kinetic}. Hence, when the walker is not in a localized state, it behaves in a similar manner as a standard random walk despite memory. This property agrees with the study of the case $\gamma=0$ and a finite memory range, where the walker diffuses normally at late times, albeit with a rescaled diffusion coefficient \cite{boyer2014solvable}. 

It follows from the analysis of Figs. \ref{fig:lifetime}a-b that intermittent localization is characterized by a finite mean localization time $\langle t_{loc}\rangle=\tau_0$, whereas the mean de-localization time $\langle t_{deloc}\rangle=\int_0^{\infty}dt\ t\rho(t)$ diverges. The latter property is a consequence of the inequality $\alpha<2$. Hence very long trajectories will be dominated by very long diffusive segments of diverging durations interspersed by relatively short localized periods, making the overall time-averaged occupation probability of the origin vanish asymptotically (when one does not distinguish between the two phases of motion). In this sense, the localized phase is \lq\lq metastable".

\begin{figure}[t]
    	\begin{center}
            \subfloat[]{%
            \includegraphics[width = 0.45\textwidth]{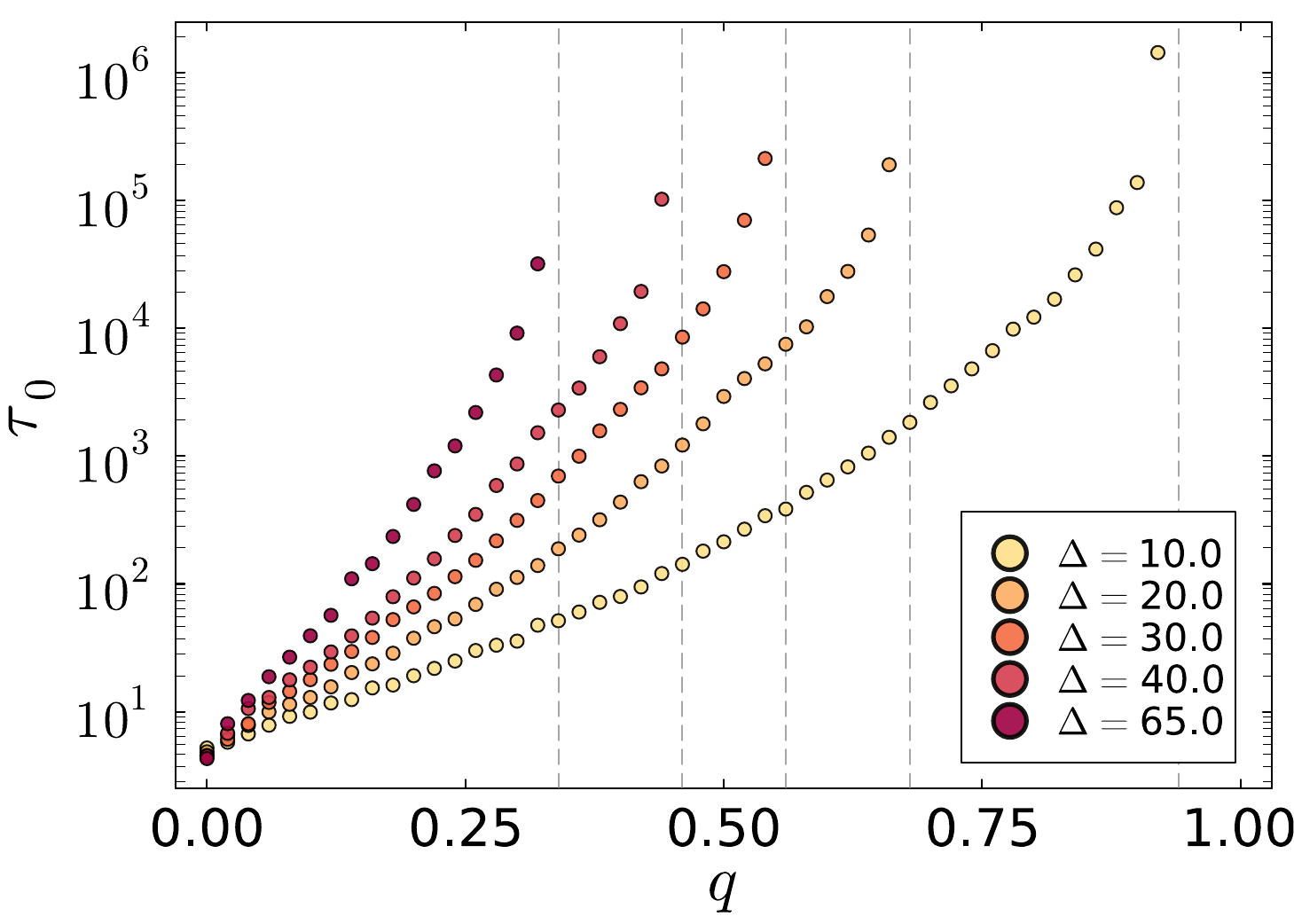}
            \label{fig:tau0_loc_exp_kernel}%
            } \hfil
            \subfloat[]{%
            \includegraphics[width = 0.45\textwidth]{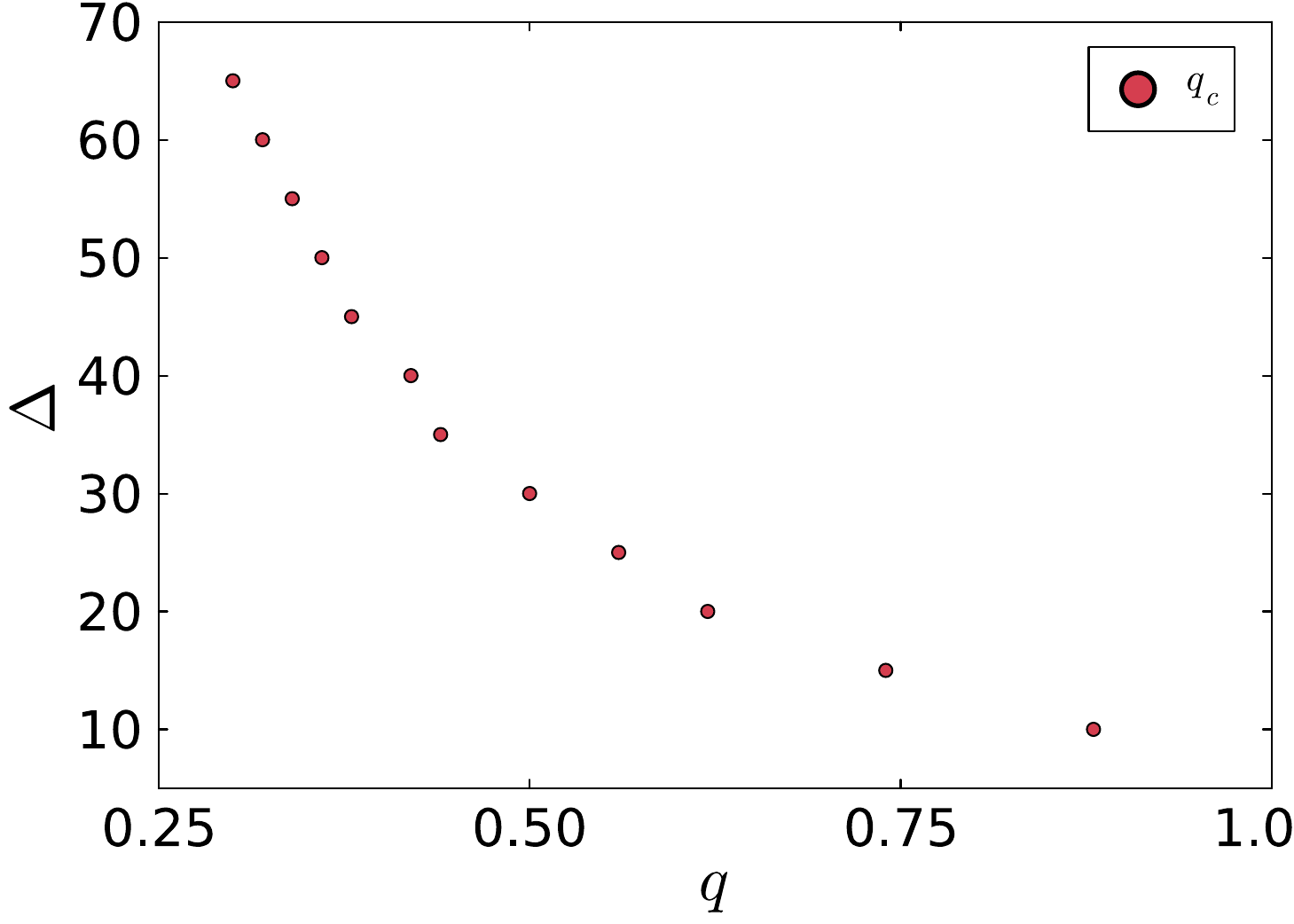}
            }
	\end{center}
    \caption{(a) Mean lifetime of the localized periods, $\tau_0$, vs. $q$. Symbols represent the inverse slope of the fits as in Fig. \ref{fig:lifetime}a for different values of $\Delta$, and fixing $\gamma = 0.7$. The vertical lines (gray) are guides to the eye to indicate a crossover value $q_c^{(I)}$.
    (b) \lq\lq Phase diagram" in the $(q, \Delta)$ space. Symbols represent the values of $q_c^{(I)}(\Delta)$, or equivalently $\Delta_c^{(I)}(q)$, below which localization is clearly intermittent.}
    \label{fig:mean_lifetime_exp}
\end{figure}

The mean lifetime $\tau_0$ of the localized periods is determined by fitting the numerical distributions of Fig. \ref{fig:lifetime}a with Eq. (\ref{plife}). Figure \ref{fig:mean_lifetime_exp}a shows the variations of $\tau_0$ vs. $q$ at fixed $\gamma=0.7$, for several choices of $\Delta$. Interestingly, $\tau_0$ sharply increases with $q$ and becomes larger than $\Delta$ by orders of magnitude. 
The simulations suggest two possible scenarios: intermittent localization turns into full localization as $q$ crosses a critical value $q_c^{(I)}$ where the mean localization time diverges; or, the mean localization time is always finite but becomes extremely large for $q$ larger than a crossover value $q_c^{(I)}$. We shall come back to these scenarios in Section \ref{sec:scenarios}. Figure \ref{fig:lifetime}b displays the 'phase diagram' showing the crossover region with dots for $\delta=0.7$: above $\Delta_c^{(I)}(q)$, the walker is localized in practice; below this value, it is intermittently localized or diffusive. 


\subsection{Power-law decaying memory with $1<\beta<2$}\label{sec:1beta2}

\begin{figure}[t]
    	\begin{center}
            \subfloat[]{%
            \includegraphics[width = 0.3\textwidth]{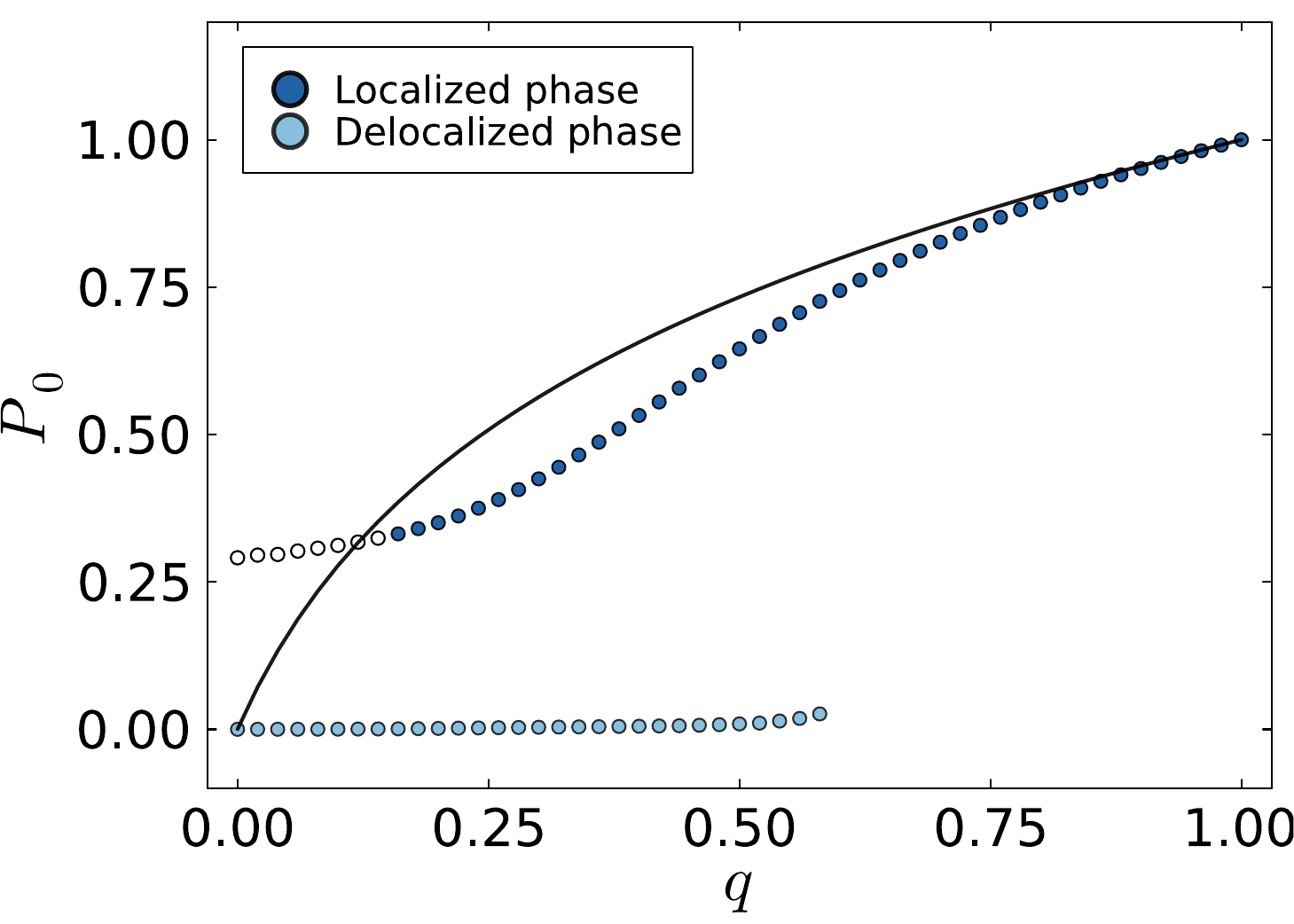}
            } \hfil
            \subfloat[]{%
            \includegraphics[width = 0.3\textwidth]{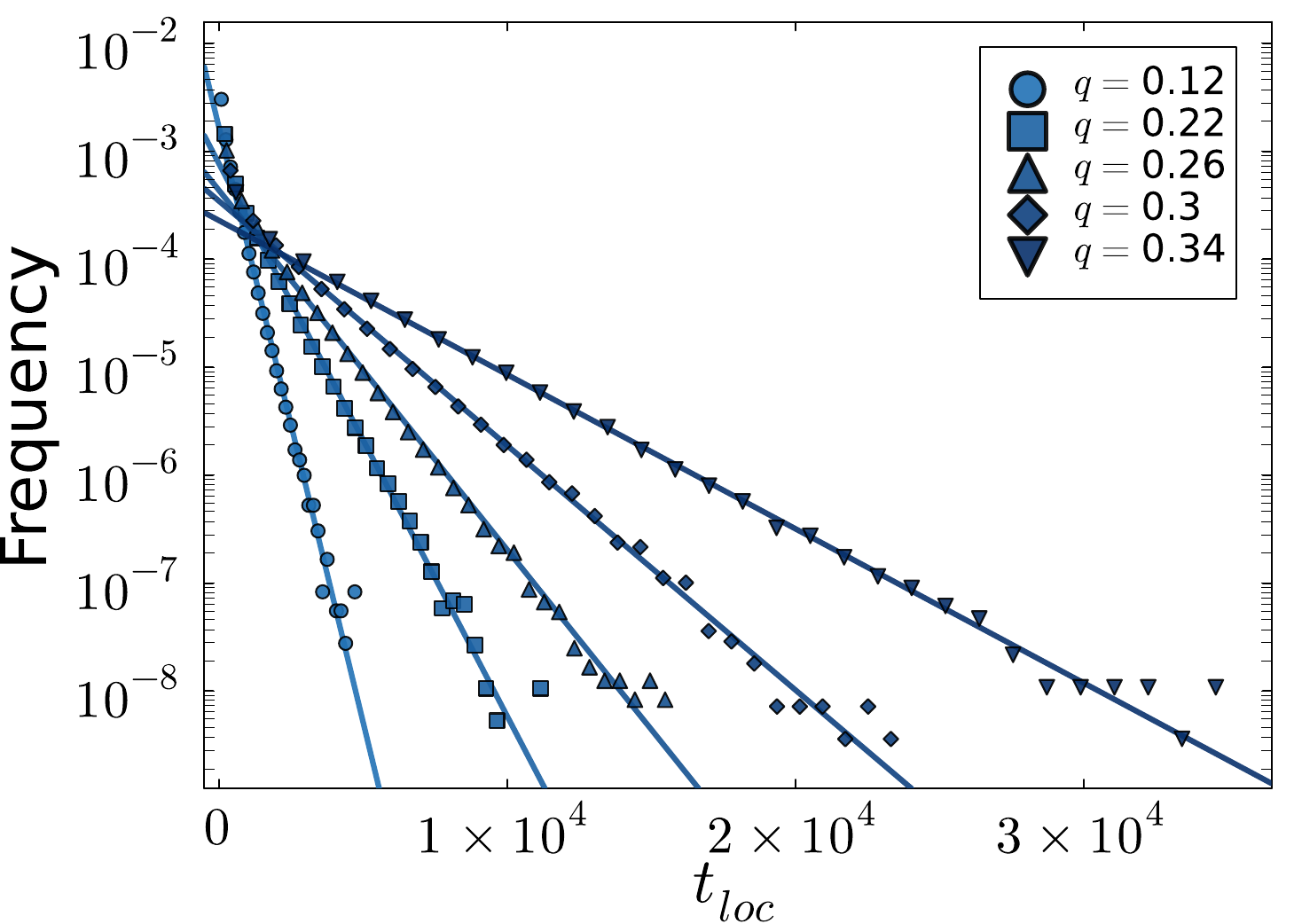}
            }  \hfil
            \subfloat[]{%
            \includegraphics[width = 0.3\textwidth]{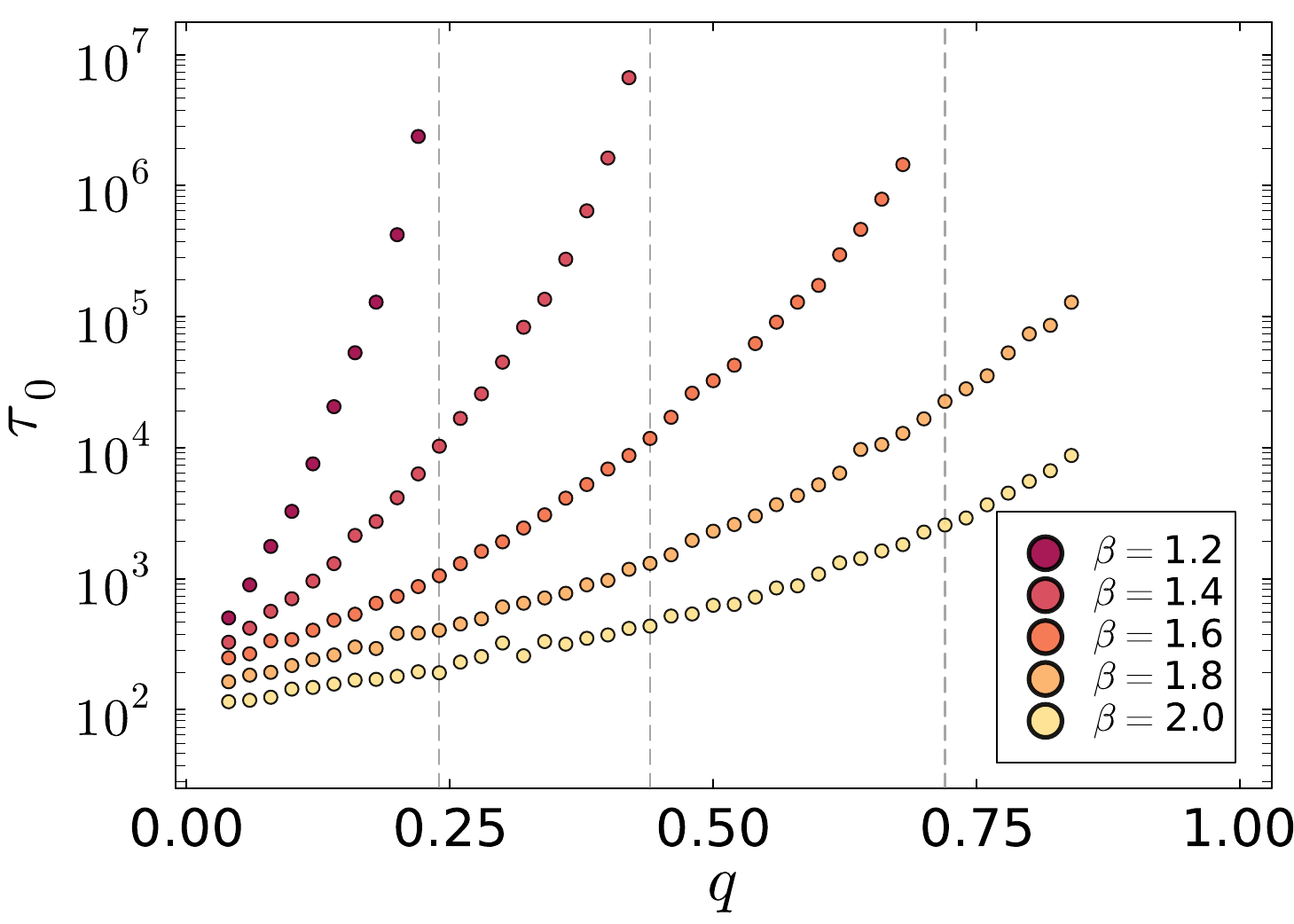}
            \label{fig:tau0_loc_powerlaw_kernel}%
            }
	\end{center}
    \caption{(a) Occupation probability of the origin in each phase. Symbols are given by Monte Carlo simulations ($t = 10^6$) for $\gamma = 0.7$ and $\beta = 1.6$, and the solid (black) line is given by Eq. (\ref{P0}).  Distributions of the lifetime $t_{loc}$ of localized periods. Symbols are given by Monte Carlo simulations ($t = 10^6$) corresponding to different values of $q$, for $\gamma = 0.7$ and $\beta = 1.6$. The solid (dark blue) lines are fits with Eq. (\ref{plife}).  (c) Mean lifetime of the localized periods, $\tau_0$, vs. $q$. Symbols represent the inverse slope of the fits as in (b) for different values of $\beta \in (1,2)$, and $\gamma = 0.7$. The vertical lines (gray) are rough estimate of the critical or crossover value $q_c^{(I)}$.}
    \label{fig:powerlawP0}
\end{figure}

Another situation of relatively fast forgetting corresponds to the kernel (\ref{pipowerlaw}) in the range $1<\beta<2$. The mean forgetting time $\langle \tau\rangle$ in Eq. (\ref{forgettime}) is still infinite in that case, but the effective number of remembered visits $C(t)$ in Eq. (\ref{C}) saturates to a finite value at large $t$. Based on the numerical results obtained by varying $\beta$ within this interval, we speculate that this case is qualitatively similar to the exponential kernel of the previous section. We were not able to observe fully localized solutions in this range, contrary to the case $0\le \beta\le 1$. As exemplified by the numerical results in Figs. \ref{fig:powerlawP0}a-b-c for $\beta=1.6$, the system exhibits intermittent localization again with an occupation probability $P_0$ for the localized phase quite close to the perfect case given Eq. (\ref{P0}). Meanwhile, the duration of the localization periods is still distributed exponentially and the mean localization time $\tau_0$ sharply increases with $q$.

\section{Linear stability analysis of the localized solution}\label{sec:stability}

In this Section, we attempt to explain some of these results by studying Eq. (\ref{ME}) beyond the steady state regime. We assume that the positions $X_{t'}$ and $X_t$ are still weakly correlated and
use the factorization in Eq. (\ref{decoupling})-(\ref{decouplingb}), but keeping the time dependence in the occupation probabilities. The master equation (\ref{ME}) becomes
\begin{eqnarray}
\label{ME2}
P_n(t+1)&\approx&\frac{1-q}{2}\left[P_{n+1}(t)+ P_{n-1}(t)\right]
+\gamma P_0(t)\left[\delta_{n,0}-\frac{1-q}{2}(\delta_{n,1} +\delta_{n,-1})\right]
\\
&+&q\sum_{t'=0}^{t}\pi_{t,t'} P_n(t')-q\gamma \sum_{t'=0}^{t}\pi_{t,t'} P_n(t')P_0(t)\,.\nonumber
\end{eqnarray}
Let us assume that $P_n(t)$ has converged to the perfect localized state $P_n^{(st)}$ given by Eq. (\ref{Pn}). At some time $t_0$, a small perturbation $g_n(t=t_0)$ is applied to the distribution and further evolves in time, 
\begin{equation}\label{pert}
P_n(t)=P_n^{(st)}+g_n(t)\,,\quad {\rm with}\quad |g_n(t)|\ll P_n^{(st)}\,,\ \ t\ge t_0\,,
\end{equation}
where $\sum_{n=-\infty}^{\infty}g_n(t)=0$ at all $t$ by normalization. Inserting 
Eq. (\ref{pert}) into Eq. (\ref{ME2}) and neglecting the term of ${\cal O}(g_n(t)^2)$ one obtains an equation for the evolution of the perturbation $g_n(t)$ at first order,
\begin{eqnarray}\label{eqg}
g_n(t+1)&=&\frac{1-q}{2}\left[g_{n+1}(t)+ g_{n-1}(t)\right]
+\gamma g_0(t)
\left[\delta_{n,0}-\frac{1-q}{2}(\delta_{n,1} +\delta_{n,-1})\right]\\
&+&q(1-\gamma P_0^{(st)})\sum_{t'=t_0}^{t}\pi_{t,t'} g_n(t')-q\gamma  P_n^{(st)}g_0(t)\,.\nonumber
\end{eqnarray}
Let us look for solutions taking the form of separation of variables, $g_n(t)=h(t)f_n$, and substitute this ansatz into Eq. (\ref{eqg}). Dividing by $h(t)f_n$ on each side, separating the time-dependent part from the space-dependent one and setting them equal to a constant $\lambda$ independent of $n$ and $t$, we obtain
\begin{eqnarray}
&&h(t+1)=\lambda h(t)+q\left(1-\gamma P_0^{(st)}\right)\sum_{t'=t_0}^t\pi_{t,t'}h(t') \label{eqh}\\
&&f_{n+1}+f_{n-1}-\frac{2\lambda}{1-q}f_n=\gamma f_0\left[\delta_{n,1}+\delta_{n,-1}+\frac{2}{1-q}\left(-\delta_{n,0}+qP_n^{(st)}\right) \right].\label{eqfn}
\end{eqnarray} 
Clearly, the spatial part in Eq. (\ref{eqfn}) is independent of the memory kernel $\pi_{t,t'}$, which appears only in the time-dependent part (\ref{eqh}).
By taking the sum of Eq. (\ref{eqfn}), one checks that $\sum_{n=-\infty}^{\infty}f_n=0$ as required. The strategy is to obtain the set of eigenvalues $\{\lambda\}$ by solving Eq. (\ref{eqfn}) for the spatial part $f_n$, asking that the solutions do not diverge at $|n|\to\infty$. The eigenvalue $\lambda$ can then be replaced into Eq. (\ref{eqh}) and the behavior of $h(t)$ at large $t$ is deduced by iteration given a kernel $\pi_{t,t'}$. If $h(t)$ decreases to zero, the mode is stable; if $h(t)$ increases, it is unstable. Due to the non-local operator in time, $h(t)$ differs from an exponential in principle.
An arbitrary perturbation can then be decomposed into a linear superposition of the eigenmodes that we denote as
\begin{equation}\label{sumpert}
g_n(t)=\sum_{\lambda} a_{\lambda} h^{(\lambda)}(t)f_n^{(\lambda)},
\end{equation}
where the amplitudes $a_{\lambda}$ depend on the initial condition, which is such that $P_n^{(st)}+g_n(t=t_0)\ge 0$ for all $n$.
In most of the following, we drop the indices $(\lambda)$ for brevity in the notation, and we set $h(t=0)=1$ by convention.

\subsection{Space-dependent part}

Setting $n=0$, $1$ and $-1$ in Eq. (\ref{eqfn}), one obtains the following relations, respectively,
\begin{eqnarray}
&&f_{-1}+f_1=2\left[\lambda'+\frac{\gamma}{1-q}(qP_0^{(st)}-1)\right]f_0\label{f1a}\\
&&f_2=2\lambda'f_1+\left(\gamma-1 +\frac{2q\gamma}{1-q} P_1^{(st)}\right)f_0\label{f2a}\\
&&f_{-2}=2\lambda'f_{-1}+\left(\gamma-1 +\frac{2q\gamma}{1-q} P_1^{(st)}\right)f_0\,,
\label{fm1a}
\end{eqnarray}
with 
\begin{equation}\label{lambdap}
\lambda'=\frac{\lambda}{1-q}\,,
\end{equation}
and where $P_n^{(st)}$ is known from Eqs. (\ref{P0})-(\ref{Pn}). For $n\ge 2$ or $n\le -2$, Eq. (\ref{eqfn}) reduces to
\begin{equation}\label{fnge2}
f_{n+1}+f_{n-1}-2\lambda'f_n=\frac{2q\gamma P_n^{(st)}}{1-q}f_0\,.
\end{equation}
Due to the symmetry $P_{-n}^{(st)}=P_{n}^{(st)}$, we focus on the positive integers $n$.

\subsubsection{Antisymmetric solutions} 
A first kind of solutions, denoted as $f_n^{(a)}$, are anti-symmetric and characterized by $f^{(a)}_0=0$. In this case, Eq. (\ref{eqfn}) is homogeneous for all $n$ and reads
\begin{equation}
f^{(a)}_{n+1}+f^{(a)}_{n-1}-2\lambda'f^{(a)}_n=0\,.
\end{equation}
One deduces $f^{(a)}_n=b r_1^n+cr_2^n$ (with $c=-b$, to enforce the condition $f^{(a)}_0=0$) where $r_1$ and $r_2$ are the root of the polynomial $r^2-2\lambda'r+1=0$, or
\begin{equation}\label{r1r2}
r_1=\lambda'-\sqrt{\lambda'^2-1},\quad r_2=r_1^{-1}=\lambda'+\sqrt{\lambda'^2-1}\,.
\end{equation}
If $\lambda'>1$ then $r_1$ and $r_2$ are reals with $r_1<1$ and $r_2>1$. Thus $f_n^{(a)}$ diverges as $r_2^{n}$ when $n\to+\infty$, and as $r_1^{-|n|}=r_2^{|n|}$ when $n\to-\infty$. Similarly for $\lambda'<-1$. These solutions are not acceptable, therefore $|\lambda'|$ must be $\le1$. Setting $\lambda'=\cos k$ with $k\in[0,\pi]$, the roots in Eq. (\ref{r1r2})
become $r_1=e^{-ik}$ and $r_2=e^{ik}$. We conclude that the anti-symmetric part of the spectrum is given by 
\begin{eqnarray}\label{scatt}
&&f_n^{(a)}=\sin(kn)\label{scattfn}\\
&&\lambda=(1-q)\cos k\,,
\end{eqnarray}
which corresponds to scattering states, or spatially extended states.

\subsubsection{Symmetric solutions} 

The second type of solutions are symmetric, $f_{-n}=f_{n}$, and denoted as $f_n^{(s)}$. Eq. (\ref{eqfn}) must be considered in full with a non-zero $f_0^{(s)}$ that can be set to $1$ (or some normalizing factor) without restricting generality.
In this case, Eq. (\ref{f1a}) becomes
\begin{equation}
f_1^{(s)}=\left[\lambda'+\frac{\gamma}{1-q}(qP_0^{(st)}-1)\right]f_0^{(s)}\label{f1b}\,.
\end{equation}
Hence, given the sole $f_0^{(s)}$, the next coefficient $f_1^{(s)}$ is deduced along with $f_2^{(s)}$ through Eq. (\ref{f2a}), and all the remaining $f_n^{(s)}$ with $n>2$ by iterating Eq. (\ref{fnge2}). The coefficients with $n<0$ follow by symmetry.

To solve Eq. (\ref{fnge2}) for $f_n^{(s)}$ with $n=3,4,...$ given the \lq\lq initial condition" $f_1^{(s)}$ and $f_2^{(s)}$, let us calculate the generating function
\begin{equation}
\widetilde{f}^{(s)}(\epsilon)=\sum_{n=1}^{\infty} \epsilon^nf_n^{(s)}\,.
\end{equation}
The poles of an arbitrary function $\widetilde{f}^{(s)}(\epsilon)$ in the complex plane, at $\epsilon=\{\epsilon_1,\epsilon_2,...\}$, correspond to exponential tails $f_n^{(s)}\sim c_1\epsilon_1^{-n}+c_2\epsilon_2^{-n}+...$
for $n>0$ large enough, where the coefficients $c_1,c_2,...$  depend on $(q,\gamma,\lambda')$. Assume that there is a pole, e.g., $\epsilon_1$, with $|\epsilon_1|<1$. To ensure that $f_n^{(s)}$ does not diverge at large $n$, this pole must be canceled by imposing $c_1=0$. Another type of divergence can be caused by the presence of a double pole $\epsilon_1=\epsilon_2=1$ (implying that $f_n^{(s)}$ grows as $n$ at large $n$). In this case, one needs to impose $c_1=0$ as well. Therefore,
solving $c_1(\lambda')=0$ gives the sought eigenvalue $\lambda'$ as a function of $q$ and $\gamma$, and finally $\lambda$ via Eq. (\ref{lambdap}).

Multiplying Eq. (\ref{fnge2}) by $\epsilon^n$, summing over $n=2,3,...$ and using the identity  $\epsilon+\epsilon^{-1}-2\lambda'=(\epsilon-r_1)(\epsilon-r_2)/\epsilon$ yields after some straightforward algebra and using the relation (\ref{f2a}),
\begin{equation}\label{ftilde}
\widetilde{f}^{(s)}(\epsilon)=\frac{\epsilon R(\epsilon,\lambda')}{(\epsilon-r_1)(\epsilon-r_2)}\,,
\end{equation}
with 
\begin{equation}\label{R}
R(\epsilon,\lambda')=\left[\frac{2q\gamma \widetilde{P}^{(st)}(\epsilon)}{1-q}+\epsilon(\gamma-1)\right]f_0^{(s)} +f_1^{(s)},
\end{equation}
where $f_1^{(s)}$ is given by Eq. (\ref{f1b}), and
\begin{equation}\label{Ptilde}
\widetilde{P}^{(st)}(\epsilon)=\sum_{n=1}^{\infty}\epsilon^n P_n^{(st)}=\frac{(1-\gamma)\epsilon P_0^{(st)}}{a-\epsilon}\,,
\end{equation}
from the known solution (\ref{Pn}).
Recall that $(r_1,r_2)$ depend on $\lambda'$ and are given by Eq. (\ref{r1r2}). 
There are two cases:

\hspace{0.5cm}$\bullet$ Case $|\lambda'|<1$. In this case, we can set
\begin{equation}\label{lp}
\lambda'=\cos k\,,
\end{equation}
with $0<k<\pi$. This means that $r_1,r_2$ are complex conjugates with $|r_1|=|r_2|=1$ like in the anti-symmetric solution. None of the 3 poles $\{r_1,r_2,a\}$ in Eq. (\ref{ftilde})-(\ref{Ptilde}) has a modulus smaller than 1 [$a>1$ from Eq. (\ref{a})], and a double pole does not exist. Therefore, no pole needs to be canceled. For $n>0$ and large enough, the eigenfunctions are of the form 
\begin{equation}\label{scattsym}
f^{(s)}_n\simeq 2{\cal R}e \left[c_1\, e^{-ikn}\right]+c_3\, a^{-n}\,,
\end{equation}
where we have used the fact that $c_2$ is the complex conjugate of $c_1$.

\hspace{0.5cm}$\bullet$ Case $\lambda'\ge 1$. In this case $r_{1,2}$ are reals, with $r_1<1$ and $r_2>1$, or $r_1=r_2=1$ (double pole). As explained above, one must cancel the pole $r_1$ in Eq. (\ref{ftilde}), which is done by imposing $R(\epsilon=r_1,\lambda')=0$. The eigenvalue is thus a root of the equation
\begin{equation}\label{polecancel}
R\left(\epsilon=\lambda'-\sqrt{\lambda'^2-1},\lambda'\right)=0\,.
\end{equation}
It is easy to check from Eq. (\ref{R}) that $R(1,1)=0$, independently of $(q,\gamma)$. This can be seen after a short calculation by substituting $f_1^{(s)}$ by
Eq. (\ref{f1b}) and using the identity $\widetilde{P}(\epsilon=1)=(1-P_0^{(st)})/2$ due to the normalization of $P_n^{(st)}$. Therefore, Eq. (\ref{polecancel}) has a remarkably simple solution,
\begin{equation}\label{lmax0}
\lambda'=1\, ,\quad{\rm or}\quad \lambda=1-q\,,
\end{equation}
for all  $0<q<1$ and $0<\gamma<1$. With the help of the software Mathematica\textsuperscript{\tiny\textregistered}, one can check graphically with many pairs of parameters $(q,\gamma)$ that Eq. (\ref{polecancel}) does not admit other real solutions.
Hence, the solution $\lambda'=1$ is unique, independent of $(q,\gamma)$, and implies that $r_1=r_2=1$ from Eq. (\ref{r1r2}). In other words,
Eq. (\ref{ftilde}) has a double pole and Eq. (\ref{polecancel}) guarantees that it is canceled. In this marginal case, corresponding to $k\to 0$ in Eq. (\ref{lp}), the eigenstate is a bound state since only the exponentially decaying mode remains, or
\begin{equation}\label{boundstate}
f^{(s)}_n|_{\lambda=1-q}\simeq c_3\,a^{-|n|}\,,
\end{equation}
at large $|n|$.

In summary, the eigenfuntions $f_n$ are given by Eqs. (\ref{scattfn}) and (\ref{scattsym}), associated to the degenerate eigenvalue $\lambda=(1-q)\cos k$ with $k\neq0$, and by Eq. (\ref{boundstate}), associated to $\lambda=1-q$. The prefactors $c_1$ and $c_3$ depend on the parameters $(q,\gamma)$ and $k$. Notice that these modes are not necessarily orthogonal.

\subsection{Time-dependent part}\label{sec:genstab}

The analysis of the spatial part has allowed us to find the largest eigenvalue (\ref{lmax0}) of the perturbation spectrum,
\begin{equation}\label{lmax}
\lambda_{max}=1-q\,.
\end{equation}
In the temporal part, $\lambda$ only appears in the first term of the rhs of Eq. (\ref{eqh}) for the evolution of $h(t)$, which involves the kernel $\pi_{t,t'}$. We expect the amplitude to grow
with time if $\lambda$ is large enough, say, above a threshold value denoted as $\lambda_c^{(I)}$, and to decay to 0 if $\lambda<\lambda_c^{(I)}$. The largest eigenvalue $\lambda_{max}$ corresponds to the most \lq\lq dangerous" mode, which dominates the dynamics at late times by growing the fastest or decaying the slowest.
Heuristically, the two types of asymptotic behaviors are separated by the marginal case at $\lambda=\lambda_c$ where $h(t)\to h^*$ at late times, with $h^*$ a constant. Substituting $h(t)$ by a constant in Eq. (\ref{eqh}) and taking $t\to\infty$, one obtains
\begin{equation}
\label{lc0}
\lambda_c^{(I)}=1-q\left(1-\gamma P_0^{(st)}\right)\lim_{t\to\infty}\sum_{t'=t_0}^{t}\pi_{t,t'}\,.
\end{equation}
For memory kernels of the form (\ref{pi}), where $F(\tau)$ is assumed to decay or to remain constant, the initial times in the interval $t'\in[0,t_0]$ contribute very little at large $t$ to the distribution $\pi_{t,t'}$. Hence, the sum in Eq. (\ref{lc0}) is equal to unity due to normalization and the equation simplifies to
\begin{equation}
\label{lc}
\lambda_c^{(I)}=1-q\left(1-\gamma P_0^{(st)}\right)=(1-q)(1+u)\,,
\end{equation}
where we recall that $u$ is given by Eq. (\ref{u}).
With Eq. (\ref{lc}), we reach two important conclusions:

\hspace{0.5cm}$\bullet$ The instability threshold $\lambda_c^{(I)}$ is independent of the memory kernel and only involves $(q,\gamma)$.

\hspace{0.5cm}$\bullet$ The inequality $\lambda_{max}<\lambda_c^{(I)}$ always holds since $u>0$, therefore the most dangerous perturbation relaxes to zero and all the localized solutions are stable. This property is also independent of the memory kernel.

These conclusions agree with some of our results. In Section \ref{sec:slowmem}, the fully localized states appeared to be stable attractors of the dynamics in the simulations, for sufficiently long-range memory kernels (see Fig. \ref{fig:powerlaw}b). This is actually the regime where the de-correlation approximation used in the stability analysis is expected to be most accurate and valid.

Unfortunately, the same conclusion predicts that localization in short-range memory systems should be stable, too, which is inconsistent with the observation of intermittent localization. Therefore, the theory is unable to describe the intermittent localization phenomenon, for which one could have, for instance, at least one unstable mode or $\lambda_{max}>\lambda_c$. This suggests that during the evolution of localization toward diffusive states, temporal correlations start to play a crucial role and must be taken into account.

\section{Relaxation dynamics toward the localized state for $0\le\beta\le1$.}\label{sec:rel} 
Let us return to the case of the power-law memory kernel (\ref{pipowerlaw}) with $0\le\beta\le1$, a regime where our linear stability analysis holds in principle. 

\subsection{Relaxation of the slowest mode}\label{sec:rel_lmax}
The way the amplitude of a perturbation relaxes to zero does depend on the memory kernel. We can apply the results of the stability analysis to the study of the relaxation of the leading eigenmode with $\lambda=\lambda_{max}=1-q$, which corresponds to Eq. (\ref{lp}) with $k=0$. At large $t$, we
approximate the sums by integrals and time is taken as continuous; in addition $h(t+1)-h(t)$ is replaced by the time-derivative $\dot{h}(t)$. Equation (\ref{eqh}) with $\lambda=1-q$ and $t_0=0$ becomes
\begin{equation}\label{conth}
    \dot{h}(t)\approx-qh(t)+q\left(1-\gamma P_0^{(st)}\right)\frac{1-\beta}{t^{1-\beta}}\int_0^{t}dt'\frac{h(t')}{(1+t-t')^{\beta}}\,.
\end{equation}
We define the operator,
\begin{equation}\label{opermem}
{\cal F}[h(t)]=\int_0^tdt'\frac{h(t')}{(1+t-t')^{\beta}}\,,
\end{equation}
and assume that the approach toward the steady state is non-exponential, and takes the form of an inverse power-law. We thus make the ansatz, 
\begin{equation}\label{ansatz1}
h(t)= t^{\nu}\quad {\rm with}\quad \nu<0 
\end{equation}
at late times, where the exponent $\nu$ is unknown. 
Following a similar method as in \cite{boyer2014solvable}, we substitute $h(t')$ by the form (\ref{ansatz1}) over the whole time domain, choose a positive $\varepsilon\ll 1$ and a large $t$ such that $\varepsilon t\gg1$. We then make the change $\tau=t-t'$ and decompose,
\begin{eqnarray}
{\cal F}[t^{\nu}]&=&t^{\nu}\left[\int_0^{\varepsilon t}d\tau\frac{(1-\tau/t)^{\nu}}{(1+\tau)^{\beta}} +
\int_{\varepsilon t}^t d\tau\frac{(1-\tau/t)^{\nu}}{(1+\tau)^{\beta}}
\right]\nonumber\\
&\simeq& t^{\nu}\left[\int_0^{\varepsilon t}d\tau\frac{1}{(1+\tau)^{\beta}} 
-\frac{\nu}{t} \int_0^{\varepsilon t}d\tau\frac{\tau}{(1+\tau)^{\beta}} 
+\int_{\varepsilon t}^t d\tau\frac{(1-\tau/t)^{\nu}}{(1+\tau)^{\beta}}
\right]\,.\label{decompF}
\end{eqnarray}
The first and second integrals in the rhs of Eq. (\ref{decompF}) are ${\cal O}\left( (\varepsilon t)^{1-\beta}\right)$ and  ${\cal O}\left(\varepsilon (\varepsilon t)^{1-\beta}\right)$, respectively, and can be neglected compared to the third term, which is ${\cal O}\left( t^{1-\beta}\right)$. To see this, let us  replace $(1+\tau)^{\beta}$ by $\tau^{\beta}$ in the denominator of the third term (since $\epsilon t\gg1$), 
\begin{equation}\label{3rd}
\int_{\varepsilon t}^t d\tau\frac{(1-\tau/t)^{\nu}}{(1+\tau)^{\beta}}\simeq
t^{1-\beta}\int_{\varepsilon}^1du\frac{(1-u)^{\nu}}{u^{\beta}}
\simeq
t^{1-\beta}\int_{0}^1du\frac{(1-u)^{\nu}}{u^{\beta}}\,,
\end{equation}
where the last equality follows from the condition $\beta<1$. For the last integral in Eq. (\ref{3rd}) to be finite, we need to assume $\nu>-1$ (to be checked below). We use the identities $\int_0^{1}du\ u^{z-1}(1-u)^{w-1}=\Gamma(z)\Gamma(w)/\Gamma(z+w)$ and $z\Gamma(z)=\Gamma(z+1)$, where $\Gamma(z)$ is the Gamma Function \cite{AS_book}, and 
substitute Eqs. (\ref{decompF})-(\ref{3rd}) into Eq. (\ref{conth}). Neglecting $\dot{h}(t)$ compared to $h(t)$, we obtain an equation independent of time for the unknown exponent $\nu$,
\begin{equation}\label{nu}
1=\left(1-\gamma P_0^{(st)}\right)\frac{\Gamma(2-\beta)\Gamma(1+\nu)}{\Gamma(2+\nu-\beta)}\,.
\end{equation}
We analyze the different cases below.

\hspace{0.5cm} {\bf $\bullet$ Case $\beta=0$ :} When memory is perfect, the solution of Eq. (\ref{nu}) is simply $\nu_{\beta=0}=-\gamma P_0^{(st)}$, which belongs to the interval $(-1,0)$, as assumed. Hence,
\begin{equation}
h(t)\sim t^{-\gamma P_0^{(st)}}\to0\quad {\rm as}\ t\to\infty\,.
\end{equation}
Hence, the amplitude of the most dangerous eigenmode relaxes to $0$ at large times, confirming the stability of the localized solution found in Section \ref{sec:genstab}. The relaxation becomes very sluggish ($\nu_{\beta=0}\to0^-$) when $\gamma$ or $q$ are small (i.e., $P_0^{(st)}$ small). In the curve $\gamma=0.4$ of Fig. \ref{fig:powerlaw}b, one actually checks that the relaxation toward the steady state solution becomes slower as $q$ decreases (see, e.g., the red circles).

\hspace{0.5cm} {\bf $\bullet$ Case $0<\beta\ll 1$ :} A first order expansion of Eq. (\ref{nu}) in $\beta$ gives
\begin{equation}
\nu=-\gamma P_0^{(st)}\left(1+\beta\frac{\int_0^1du\,(1-u)^{-\gamma P_0^{(st)}-1}\,u\ln u}{\int_0^1du\,(1-u)^{-\gamma P_0^{(st)}-1}\,\ln (1-u)}+{\cal O}(\beta^2)\right)<\nu_{\beta=0}\,.
\end{equation}
Therefore, when memory slowly decays with time, the relaxation towards the localized state is {\em faster} compared to the perfect memory case $\beta=0$. 

\hspace{0.5cm} {\bf $\bullet$ Case $\beta$ close to $1$ :} Let us set $\beta=1-\varepsilon_1$ and assume $\nu=-1+\varepsilon_2$, with   $0<\varepsilon_1\ll 1$ and $0<\varepsilon_2\ll 1$. Using $\Gamma(1)=1$ and the expansion $\Gamma(z)\simeq 1/z$ at small $z$ in  Eq. (\ref{nu}), one obtains  $\varepsilon_2=\varepsilon_1/[(1-\gamma P_0^{(st)})^{-1}-1]$, or,
\begin{equation}
\nu=-1+\frac{1-\beta}{\left(1-\gamma P_0^{(st)}\right)^{-1}-1}+{\cal O}((1-\beta)^2),\quad\beta\to1^{-}.
\end{equation}
Hence, when $\beta$ approaches 1 from below, the relaxation at large times is much faster than for small values of $\beta$ and it follows a generic form approximately given by $h(t)\sim t^{-1}$, independently of $(q,\gamma)$. As shown below, this result suggests that the fastest relaxation toward the steady state, among the values of $\beta$ in the interval $[0,1]$, is achieved when $\beta$ is exactly $1$. 

\hspace{0.5cm} {\bf $\bullet$ Boundary case $\beta=1$ :} Eq. (\ref{nu}) no longer admits a solution and does not describe this case. After replacing the sum $\sum_{t'=0}^{t}$ in Eq. (\ref{pi}) by $\int_0^t dt'$,
the memory kernel for $\beta=1$ can be approximated by $\pi_{t,t'}\simeq [\ln(1+t)]^{-1}/(1+t-t')$. Setting $\lambda=\lambda_{max}$ in the evolution equation (\ref{eqh}) of $h(t)$ yields at large $t$,
\begin{equation}\label{eqhbeta1}
h(t)\ln (t)\simeq\left(1-\gamma P_0^{(st)}\right)\int_0^{t}dt'\frac{h(t')}{1+t-t'}\,,
\end{equation}
where once again $h(t+1)-h(t)$ has been neglected compared to the other two terms. The power-law ansatz (\ref{ansatz1}) of the relaxation at late times must be modified and we now assume
\begin{equation}\label{ansatz2}
h(t)=\frac{(\ln t)^{\alpha}}{t}\, ,
\end{equation}
with $\alpha$ an exponent.
As detailed in Appendix \ref{sec:relbeta1}, the ansatz (\ref{ansatz2}) turns out to be the one that solves Eq. (\ref{eqhbeta1}) at large $t$. The exponent $\alpha$ is given by 
\begin{equation}\label{alpha}
\alpha=\frac{1}{\gamma P_0^{(st)}}-2\,.
\end{equation}
The $1/t$ convergence toward the localized state exhibits a logarithmic correction that depends on the parameters $(q,\gamma)$. If $\gamma P_0^{(st)}<1/2$, then $\alpha>0$ and the relaxation is a bit slower than for a stronger localization, i.e., $\gamma P_0^{(st)}>1/2$, where  $\alpha<0$. When $\gamma P_0^{(st)}$ is close to 0, the exponent $\alpha$ becomes very large. Since $h(t)$ is a decreasing function of time, the asymptotic regime (\ref{ansatz2}) with a large $\alpha$ is actually reached at extremely large times and is very difficult to observe numerically.

\begin{figure}
    \centering
    \includegraphics[width=0.5\linewidth]{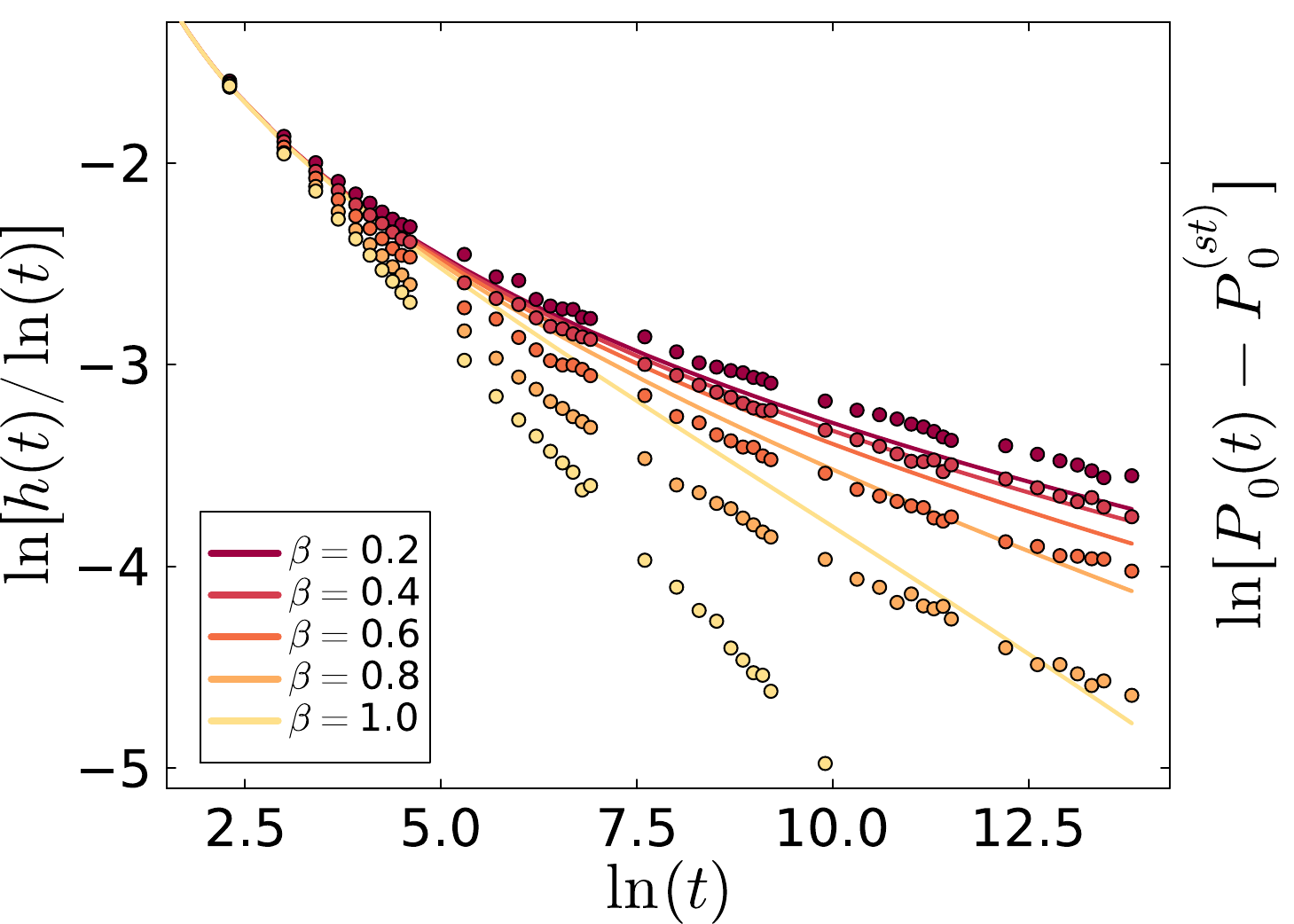}
    \caption{Relaxation toward the localized state. The theoretical amplitude $h(t)$ of the slowest mode $\lambda_{max}=1-q$ is represented (solid lines) for different values of $\beta \in[0,1]$. The reason of the division by $\ln t$ is explained in Section \ref{sec:rel_k}. The amplitudes $h(t)$ are obtained from numerically solving Eq. (\ref{eqh}) with the initial conditions $t_0 = 0$ and $h(0) = 1$ and parameters values $q = 0.1$, $\gamma = 0.4$. Symbols (same color code) represent the relaxation of the occupation probability $P_0(t)$ obtained from simulations, toward the stationary solution $P_0^{(st)}$ given by Eq. (\ref{P0}). For the boundary case $\beta=1$, the simulation value at $t=10^6$ is taken as the asymptotic one, as it is a bit below the theoretical $P_0^{(st)}$. We have multiplied the theoretical curves by a same prefactor, such that they coincide at short times with the simulations.}
    \label{fig:relaxation_time}
\end{figure}
In Figure \ref{fig:relaxation_time}, we have represented the relaxation obtained by iterating Eq. (\ref{eqh}) numerically for the leading mode $\lambda_{max}$, starting from the initial condition $h(t=0)=1$ (solid lines). As explained in the next subsection, we have displayed $h(t)/\ln t$ rather than $h(t)$ alone, as the former function includes the relaxation of the modes which are close to $\lambda_{max}$ instead of $\lambda_{max}$ alone. [Qualitatively similar results are obtained by representing $h(t)$.] The figure also shows the quantity $P_0(t)-P_0^{(st)}$, where $P_0(t)$ is obtained from numerical simulations of a walker starting at the origin at $t=0$, while $P_0^{(st)}$ is given by Eq. (\ref{P0}). 
Although the agreement is not quantitative, the theory exhibits the correct relaxation shape. And, as predicted, the simulated relaxation becomes faster as $\beta$ increases from 0 and approaches 1 (yellow curve/dots). This effect is even stronger in the simulations than in theory.

\subsection{Relaxation of the modes close to $k=0$}\label{sec:rel_k}

In the comparison with the numerical simulations of Fig. \ref{fig:relaxation_time}, we have also taken into account the eigenvalues that are close to $\lambda_{max}=1-q$ in the relaxation. As the eigenvalue spectrum $\lambda=(1-q)\cos k$ is continuous, the modes with $0<k\ll1$ significantly contribute, too, and their overall effect turns out to be a logarithmic correction to the leading power-law decay (see \cite{boyer2024power} for a similar example). For any fixed $\beta\in[0,1)$, we find that the relaxation to the stationary state is given by
\begin{equation}\label{rellog}
P_n(t)-P_n^{(st)}\sim t^{-|\nu|}/(\ln t)^{\eta},\quad {\rm with}\ \eta=1,
\end{equation}
instead of $t^{-|\nu|}$ alone.
To give an idea of the calculation, we rewrite the general evolution in Eq. (\ref{sumpert}) as
\begin{equation}\label{Pnexpand}
P_n(t)=P_n^{(st)}+\int_0^{\pi}dk\,\left( \sum_{i=a,s}a_k^{(i)}f_n^{(i)(k)}\right)h^{(k)}(t),
\end{equation}
where the dependence on $k$ has been made explicit in the functions $f_n$ and $h(t)$, while the coefficients $a^{(i)}_k$ is to be determined from the initial condition. 
Applying the ansatz $h^{(k)}(t)=t^{\nu_k}$ to Eq. (\ref{eqh}) with the eigenvalue $\lambda=(1-q)\cos k$ leads to a modified Eq. (\ref{nu}) for the exponent $\nu_k$ of that mode,
\begin{equation}\label{nuk}
\frac{1-q}{q}(1-\cos k)+1=\left(1-\gamma P_0^{(st)}\right)\frac{\Gamma(2-\beta)\Gamma(1+\nu_k)}{\Gamma(2+\nu_k-\beta)},
\end{equation}
from which we deduce, at small $k$,
\begin{equation}
\nu_k\simeq\nu- ck^2
\end{equation}
with $c$ a positive constant. Therefore,
\begin{equation}\label{relhk}
h^{(k)}(t)\simeq c'\, t^{-|\nu|}e^{-ck^2\ln t},
\end{equation}
with $c'$ depending little on $k$ [recall that $h^{(k)}(t=0)=1$ by convention]. As clear from Eq. (\ref{relhk}), at large $\ln t$ the integral in Eq. (\ref{Pnexpand}) is dominated by the small $k$ regime of a Gaussian and the integration domain can be extended to $k=\infty$. Fixing $n$ (e.g. $n=0$) we therefore need to determine the behavior of $a_k^{(a,s)}$ and $f_n^{(a,s)(k)}$ at small $k$. Due to the cancellation of the double pole, $f_n^{(s)(k)}$ is dominated by the bound state (\ref{boundstate}) which depends little on $k$ close to $k=0$, i.e.,
\begin{equation}\label{fsksmall}
f_n^{(s)(k)}\sim k^{0},
\end{equation}
while the contribution of the antisymmetric part $f_n^{(a)(k)}$ given by Eq. (\ref{scattfn}) is ${\cal O}(k)$ and therefore negligible. To calculate the coefficients $a_k^{(s)}$, one writes Eq. (\ref{eqfn}) as ${\cal L}[f]=\lambda f$, with $f$ the vector of components $f_n$, and notices that the operator ${\cal L}$ is not self-adjoint. After determining the adjoint operator ${\cal L}^{\dagger}$, one obtains the eigenstates of the latter by solving ${\cal L}^{\dagger}[\widetilde{f}]=\lambda \widetilde{f}$. The results are $\widetilde{f}^{(a)(k)}_n=\sin(kn)$ and
\begin{eqnarray}
&&\widetilde{f}^{(s)(k)}_n=\mu_1\sin(k|n|)-\sin[(|n|-1)k]\label{fts}
\end{eqnarray}
(up to prefactors)
for the antisymmetric and symmetric solutions, respectively, with $\mu_1$ a function of $(q,\gamma)$. From Eq. (\ref{fts}),  at fixed $n$, one deduces the small $k$ behavior
\begin{equation}\label{tfsksmall}
\widetilde{f}_n^{(s)(k)}\sim k.
\end{equation}
The coefficients $a_k^{(s)}$ follow from the initial perturbation $P_n(t=0)=P_n^{(st)}+g_n^{(0)}$ and from the orthogonality relation $\sum_{m=-\infty}^{\infty}\widetilde{f}_m^{(s)(k')}f_m^{(s)(k)}=\delta(k-k')$. Setting $h^{(k)}(t=0)=1$ in Eq. (\ref{Pnexpand}), multiplying this equation by $\widetilde{f}_n^{(s)(k')}$, summing over $n$ and integrating over $k'$, one obtains
\begin{equation}\label{ask}
a_k^{(s)}=\sum_{m=-\infty}^{\infty}g_m^{(0)}\widetilde{f}_m^{(s)(k)}\sim k\,,\ {\rm from\ Eq.\ }(\ref{tfsksmall}).
\end{equation}
In the last approximation of Eq. (\ref{ask}), the initial perturbation $g_m^{(0)}$ is assumed to be sufficiently localized near the origin (i.e., vanishing for $|m|\gg1$).
Gathering Eqs. (\ref{relhk}), (\ref{fsksmall}) and (\ref{ask}) into Eq. (\ref{Pnexpand}), one gets $P_n(t)-P_n^{(st)}\sim t^{-|\nu|}\int_0^{\infty} dk\,k\, e^{-ck^2\ln t}$, which yields Eq. (\ref{rellog}) after integration.

\section{Dynamical scenarios of intermittent localization}\label{sec:scenarios}

\begin{figure}[h!]
    	\begin{center}
            \subfloat[]{%
            \includegraphics[width = 0.45\textwidth]{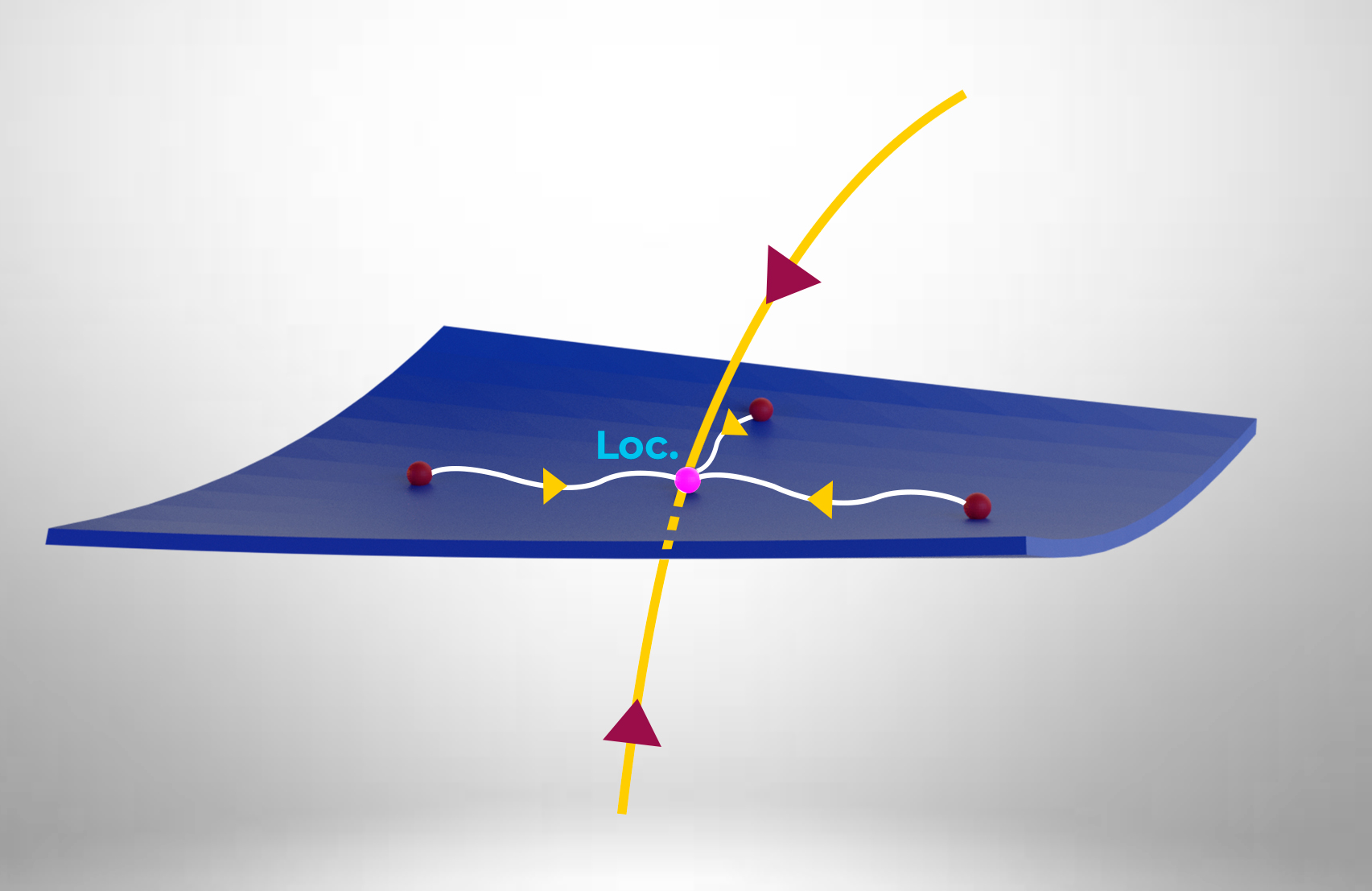}
            } \hfil
            \subfloat[]{%
            \includegraphics[width = 0.45\textwidth]{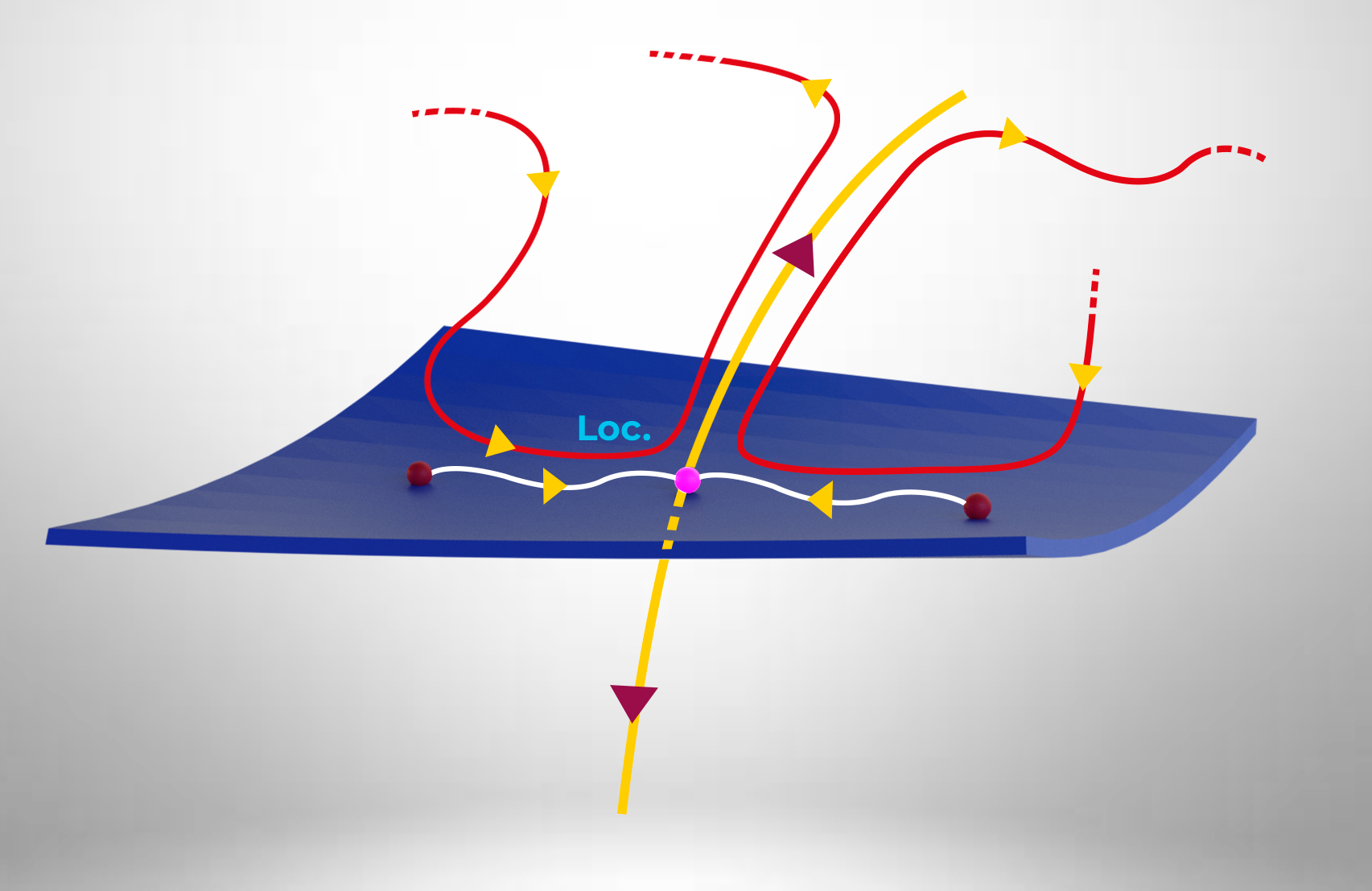}
            }
	\end{center}
    \caption{Schematic
    evolution of trajectories in the high-dimensional space of distribution functions. (a) Long range memory systems: all initial conditions are locally attracted to the fixed point (Loc.) representing the localized state in Eq. (\ref{Pn}). The one-dimensional manifold represents the eigenstate with the largest eigenvalue. (b) Short range memory systems: the mode with largest eigenvalue is now unstable. Trajectories intermittently approach the stable manifold after a possibly long excursion, are then attracted toward the fixed point and stay in its vicinity during a finite time, before being repelled into a new excursion. 
    }
    \label{fig:pheno}
\end{figure}

From a dynamical system point of view, the stability analysis of Section \ref{sec:stability} can be summarized by the phenomenological diagram of Fig. \ref{fig:pheno}a, which sketches a high-dimensional space of distribution functions. Fixing the parameters $(q,\gamma)$ and assuming that memory is sufficiently long-range ($0\le\beta\le1$), the purple fixed point represents the localized solution $P_n^{(st)}$ given by Eq. (\ref{Pn}). All trajectories/initial conditions eventually converge to this point. We have singled out the one-dimensional manifold representing the eigenmode (\ref{boundstate}) with the largest eigenvalue $\lambda_{max}$ or slowest decay.

Based on the simulation results of Sections \ref{sec:expmem} and \ref{sec:1beta2}  for the exponential and power-law cases ($1<\beta\le2$),
a possible qualitative picture for short-range memory systems is displayed in Fig. \ref{fig:pheno}b. In this representation, we have assumed that the eigenstates are stable, except the one with the largest eigenvalue, which is unstable. This situation would arise, for instance, if the generating function had a single real pole with $|\epsilon_1|< 1$ which is canceled by a sufficiently large $\lambda_{max}$, such that $\lambda_{max}>\lambda^{(I)}_c$. The stable manifold, on the other hand, must still be high-dimensional as it allows many trajectories to reach the vicinity of the localized state, although for a finite period of time due to the repulsion produced by the unstable manifold. The fixed point is essentially the same as in the stable case of Fig. \ref{fig:pheno}a.

In Fig. \ref{fig:pheno}b, when a trajectory in this distribution space moves away from the stable manifold (i.e., the particle starts to diffuse in physical space), it can wander during a long period of time before reaching a point located very close to the stable manifold again. From there, the system is attracted toward the localized state and subsequently repelled close to the unstable manifold into the next excursion. The localization time $t_{loc}$ will depend on how close the system approaches the stable manifold after an excursion: the closer it gets, the longer $t_{loc}$. The mean localization time $\tau_0$ will depend on how repulsive is the unstable manifold.

Given a sufficiently broad memory kernel $\pi_{t,t'}$, the diagrams in Figs. \ref{fig:pheno}a-b suggest two possible scenarios as the parameters $(q,\gamma)$ are varied: 

\hspace{0.5cm}  (1) The system does not exhibit a transition between the cases of Fig. \ref{fig:pheno}a and Fig. \ref{fig:pheno}b for any parameter value $q$ or $\gamma$. In other words, it is either $a)$ always fully localized, or $b)$ always intermittently localized.

\hspace{0.5cm}  (2) The type of localization of a system can belong to the cases of Fig. \ref{fig:pheno}a or Fig. \ref{fig:pheno}b, depending on $(q,\gamma)$. This means that there exists a critical value $q_c^{(I)}$ (which depends on $\gamma$ and the memory kernel parameters) such that the system is intermittently localized if $q<q_c^{(I)}$ and fully localized if $q>q_c^{(I)}$. This transition is produced by a change in stability of the most dangerous eigenmode.

We have presented arguments supporting the fact that the power-law kernel (\ref{pipowerlaw}) with $0\le \beta\le 1$ belongs to the case (1), as it appears to be always localized. 

We speculate that a necessary condition for intermittent localization is that the effective number of remembered visits $C(t)$ does not grow unbounded at late times but tends to a constant. This is the case
for memory kernels that are exponential or algebraic with $\beta>1$. The numerical results clearly show that the mean localization time becomes extremely large as $q$ increases (Figs. \ref{fig:mean_lifetime_exp}a and \ref{fig:powerlawP0}c). However, at this point, we cannot determine precisely whether these examples belong to the cases (1) or (2). 

\section{Conclusions}\label{sec:concl}

Previous work has indicated the existence of localized states for certain random walks with preferential resetting to sites visited in the past \cite{FBGM2017,BFGM2019,sepehrinia2025localization}. A simple model considers an infinite $1d$ lattice, with a single impurity site representing a resource patch with a larger attractiveness than the other sites. Due to the reinforced dynamics, the walker does not diffuse away at late times but tends to an exponential stationary distribution peaked at the impurity. This emergent property can be associated to a phenomenon of spatial learning, mimicking the activity of a foraging animal that would settle down thanks to memory around a zone where resources are abundant in a scarce environment. In the present work, we have studied how such a localization is affected by imperfect memory, when the walker slowly forgets about the steps it performed further in the past. This assumption is more realistic for animals and humans and also gives rise to new phenomena. 

As a first result, we have observed that if memory decays slowly enough, such that the effective number of remembered time-steps still grows unbounded with time (albeit sub-linearly), the walker keeps localizing and its stationary density is actually the same as in the perfect memory case. Secondly, according to a de-correlation approximation, these localized solutions turn out to be linearly stable. The relaxation of small perturbations depends on the memory kernel and is not exponential in time: the leading perturbation mode relaxes to zero as an inverse power-law $t^{-|\nu|}$ with a non-trivial exponent which depends on the parameters of the model. We note that a power-law is also thought to describe the relaxation of the classic vertex reinforced random walk, in connection with urn models \cite{pemantle1999vertex}.

Thirdly, a memory loss of the form of $1/\tau$ represents a boundary case which displays interesting peculiarities: it is the memory that decays the fastest and still produce the same localization as the perfect memory case. In other words, this memory is the \lq\lq cheapest", since the effective number of remembered steps up to time $t$ only grows as $\ln t$. In addition, somehow counter-intuitively, a walker using the $1/\tau$-kernel exhibits the {\em fastest} relaxation toward the localized distribution, see Eq. (\ref{ansatz2}).  Hence, such a walker learns faster than a walker with better memory. The speed of learning measured through the exponent $|\nu|$ obtained from Eq. (\ref{nu}) increases as $\beta$, the exponent that describes the decay of memory, varies 
from $0$ to $1$. This result resonates with empirical observations in the neurosciences indicating that forgetting is essential to the brain and can actually be beneficial to learning processes \cite{markovitch1988role,wixted2005theory,hardt2013decay}. Memory kernels in $1/\tau$ have been inferred from bison foraging data \cite{merkle2014memory} and can emerge in mean-field models of interacting synapses subject to Hebbian connections and competition  \cite{luck2014slow}.  

As a fourth result, spatial learning starts to be disrupted when the memory kernel decays exponentially, or faster than $1/\tau$, i.e., when the effective number of remembered steps saturates to a constant value at late times. A new regime of intermittent localization appears, which consists of random time intervals of localization separated by random diffusive excursions that wander away from the impurity. Numerically, the duration of these localized periods is exponentially distributed whereas that of the diffusive excursions follows a power-law distribution with infinite first moment. Consequently, over a very long observation window, the walker will tend to be found mostly in the non-localized state. Nevertheless, as the parameter of memory use ($q$) increases beyond a crossover value, the mean duration of the localized periods becomes extremely large. Remarkably, during a localization period, the distribution of the walker's position is practically undistinguishable from the localized profile with perfect memory, even when the localization periods are relatively short. Unfortunately, in the intermittent localization regime, the de-correlation approximation breaks down and is not able to predict the loss of stability (or the absence) of stationary solutions.

Intermittency in low-dimensional deterministic dynamical systems is well-known to occur through a saddle-node bifurcation, beyond which an attractor fixed point no longer exists, allowing intermittent bursts of chaos \cite{pomeau1980intermittent}. Just at this bifurcation, the relaxation toward the marginal fixed point follows a power-law form in $1/t$, which is similar although not strictly identical to Eq. (\ref{ansatz2}) for our boundary case. The $1/t$ relaxation at the edge of intermittency in deterministic systems corresponds to a critical slowing down, which is generic at bifurcation points and contrasts with the usual exponential evolution of perturbations elsewhere. In our random walk model, however, the relaxation at the edge of intermittency is faster than the one of the \lq\lq regular" localized phase: the reason is that the latter phase is also critical and characterized by a power-law with a smaller exponent $|\nu|$. Hence, it is more adequate to talk here about a critical speed-up when the intermittency threshold is approached.

Many theoretical questions remain pending, such as a proper description of intermittent localization and the surprisingly simple exponential distribution of localization times. In addition, it would be instructive to investigate the role of memory decay on the properties of the monkey walk (in particular the critical speed-up effect) in dimension two or larger, or when the underlying process is a L\'evy flight. More complex, time dependent environments should also be considered.

\acknowledgements We thank  Fabiola Monserrat P\'erez Rubio
for technical support in the elaboration of Fig. \ref{fig:pheno}, and Alberto Robledo for discussions. DB acknowledges support from CONACYT (Mexico) grant Ciencia de Frontera 2019/10872.

\appendix 

\section{Random number generation}\label{sec:rndgen}

In the Monte Carlo simulations, we employed the inverse transform sampling method to generate random numbers. This method enables drawing from arbitrary distributions using uniformly distributed random numbers. A random number $u$ is first sampled from a uniform distribution over $[0,1]$, $u \sim U(0,1)$, to sample a past time $t^{\prime}$, or equivalently, a time interval $\tau = t - t^{\prime}$, from the distribution
\begin{equation}
\label{eq:dist_tau}
    \phi(\tau) = \frac{F(\tau)}{\sum_{\tau'=0}^t F(\tau')} = \frac{F(\tau)}{C(t)}\,,
\end{equation}
for $\tau \in [0,t]$, where $F(\tau)$ is the memory kernel and the dependence of $\phi(\tau)$ on $t$ is implicit. 

\subsection{Power-law kernel}
For the power-law memory kernel defined in Section \ref{sec:slowmem} as $F(\tau) =(1 + \tau)^{-\beta}$, the target distribution is
\begin{equation}
\phi(\tau)=\frac{1}{C(t) \, (1 +\tau)^{\beta}}\,,
\end{equation}
\noindent with $\beta \geq 0$. By employing a continuous-time approximation to improve computational efficiency, the normalization constant can be determined from the condition
\begin{equation}
\label{eq:normalization_considiton_powerlaw}
\int_0^t \frac{1}{C(t) (1+\tau)^{\beta}} d \tau = 1.
\end{equation}

If $\beta \neq 1$, the normalization constant is
\begin{equation}
C(t) = \frac{(1+t)^{1-\beta}-1}{1-\beta}.
\end{equation} 
A sample from the probability distribution $\phi(\tau)$ is obtained by equating the uniform random variable $u$ in $[0,1]$ to the cumulative distribution function $\int_0^{\tau}\phi(\tau')d\tau'$ and solving for $\tau$. We thus have
\begin{equation}\label{inv}
    u = \int_{0}^{\tau} \frac{(1+\tau^{\prime})^{-\beta}}{C(t)} d \tau^{\prime}\, .
\end{equation}
After integrating,
\begin{equation}
 u  = \frac{(1+\tau)^{1-\beta}-1}{(1+t)^{1-\beta}-1}\,,
\end{equation}
we obtain the random time $\tau$ as
\begin{equation}
\tau=\left [ u\left((1+t)^{1-\beta}-1\right)+1\right] ^{\frac{1}{1 -\beta}}-1\,.
\end{equation}
Since time is a discrete variable, we take the integer part $[\tau]$ of the expression above.

For the case $\beta = 1$, we have $C(t) = \ln(1+t)$ from Eq. (\ref{eq:normalization_considiton_powerlaw}), and Eq. (\ref{inv}) gives
\begin{equation}
    \tau =  (1+t)^{u} - 1\,,
\end{equation}
from which the integer part $[\tau]$ is obtained.

\subsection{Exponential decay kernel}
Similarly, for the exponential decay kernel defined in  Section \ref{sec:expmem} as
$F(\tau) = e^{-\tau/\Delta}$,
the target distribution is
\begin{equation}
\phi(\tau)=\frac{e^{-\tau / \Delta}}{C(t)}\,,
\end{equation}
with $\Delta>0$ and $C(t)$ is given by
\begin{equation}
C(t)=  \Delta\left(1-e^{-t/\Delta}\right)\,.
\end{equation}
The equality $u=\int_0^{\tau}\phi(\tau')d\tau'$ becomes 
\begin{equation}
    u =\int_0^\tau \frac{e^{-\tau^{\prime} / \Delta}}{C(t)} d \tau^{\prime} = \frac{1-e^{-\tau/\Delta}}{1-e^{-t/\Delta}}\,.
\end{equation}
Again, the continuous-time approximation is used to reduce computational cost. By isolating $\tau$ and taking its integer part we obtain
\begin{equation}
[\tau] = \left[ -\Delta \ln\left(1 + u \left( e^{-t/\Delta} -1 \right) \right) \right]\,.
\end{equation}

\section{Phase separation algorithm}\label{sec:phasesep}

For the exponential memory kernel, we implemented an algorithm to decompose a trajectory into localized and diffusive phases. This procedure partitions the trajectories into two interval types, depending on whether the walker often occupies the impurity located at the origin (algorithm \ref{alg:interval_identification}). Each interval is then classified as localized or de-localized (algorithm \ref{alg:localized_phase}) based on whether it exceeds a minimum threshold duration, $l_{loc}$ or $l_{deloc}$, respectively. The first $3 \Delta$ steps were discarded from all trajectories to eliminate initial transients.

More precisely, given a trajectory of $T$ steps,
we determine whether the walker occupies the impurity at the origin by defining a binary variable, $\chi_t$, such that $\chi_t = 1$ when $X_t=0$ and $\chi_t =0$ otherwise. From the cumulative sum $S_t$ of this binary variable, we use the algorithm \ref{alg:interval_identification} to identify two types of intervals: increasing segments, corresponding to successive steps with $X_t=0$, and constant segments, corresponding to $X_t \neq 0$. This algorithm uses the change of the slope of $S_t$ between two successive time-steps as a criterion to determine the type of interval.

After segmenting the trajectory into increasing and constant segments, we assign to each interval one of the two phases using the algorithm \ref{alg:localized_phase}. First, we iterate over consecutive increasing intervals and measure the distance between them. Note that, by construction, two consecutive increasing intervals are always separated by a constant one. Then, we classify the intervals. For this purpose, we define a minimum length for de-localized intervals, $t_{deloc}$, depending on the memory kernel. 

For the exponential memory kernel, we define $l_\text{deloc}$ in terms of the $\Delta$, the characteristic time scale of the memory range. We arbitrarily classify all constant intervals longer than $l_{deloc}= 3\Delta$ as de-localized. In contrast, for the power-law memory kernel, the parameter $\beta$ cannot be directly interpreted as a characteristic scale for the memory. To address this, we compared the curves of the time averaged $P_0$ vs. $q$ (up to $t=10^6$) between the exponential and power-law memory kernels, identified the ones that were nearly superposed and established a linear correspondence between the values of $\Delta$ and $\beta$. Each value of $\beta$ was thus mapped to a corresponding $\Delta$ value, allowing us to define the minimum length of the de-localized intervals as $l_\text{deloc} =3\Delta$.\\

If the distance between two increasing intervals exceeds $l_\text{deloc}$, both segments are assigned to the localized phase, while the separating constant interval is classified as de-localized. In contrast, if the distance is less than or equal to $l_\text{deloc}$, the pair of increasing intervals and the constant interval are combined into a single localized interval. The merged interval is re-incorporated into the iteration process, which repeats until all intervals are classified. In our implementation, we first identify the localized intervals (algorithm \ref{alg:localized_phase}), and define the de-localized intervals as their complement.  Note that the de-localized phase is composed exclusively of segments longer than $l_\text{deloc}$.

A final filtering step is applied to the localized intervals on the basis of the interval length. Since $ 1  - \gamma$ represents the probability that the walker stays only one unit of time while visiting the impurity, $1 /( 1  - \gamma)$ is the average trapping time. For this reason, we imposed that increasing intervals shorter than $3 / (1 - \gamma)$ are excluded from the localized phase, as they are not long enough to reflect the localization behavior. These intervals are not considered in the analysis of either phase.

\begin{algorithm}[H]
\caption{\textbf{Interval Identification.} Segmentation of the trajectory into increasing and constant intervals.}
\label{alg:interval_identification}
\begin{algorithmic}[1]

\Statex \textbf{Input:} 
$t_{\text{initial}}: \texttt{Int}$\text{. Lower bound of the trajectory from which interval identification is performed.}
\Statex $\hspace{1.1cm} S_t: \texttt{Array(Int)}$\text{. Cumulative sum, $S_t$, of the binary variable, $\chi_t$, representing impurity occupation at time $t$.}

\Statex \textbf{Output:} \text{increasing\_intervals: }\texttt{Array(Array(Int))}\text{. Array storing the start and end points of the increasing intervals.}

$\hspace{0.7cm}$  \text{constant\_intervals: }\texttt{Array(Array(Int))}\text{. Array storing the start and end points of the constant intervals.}

$\hspace{0.7cm}$ \text{intervals\_type: }\texttt{Array(String)}\text{. Array containing the classification (increasing or constant) for each identified segment.}

\State $\textbf{Initialization: } L, R, t_i \text{: Int64; } m \text{ :}\texttt{Float} \text{; type: }\texttt{String}$
 \State $L \gets t_{\text{initial}}$ \Comment{Left interval endpoint}
\State $R \gets t_{\text{initial}} + 1$ \Comment{Right interval endpoint}
\State $m \gets S_t[R] - S_t[L]$ \Comment{Slope between two consecutive steps}
\State $l_S \gets \text{length} (S_t)$
\State \text{increasing\_intervals} $\gets [\ ]$
\State \text{constant\_intervals} $\gets [\ ]$
\If{$m > 0$}
    \State $\text{type} \gets$ \text{``increasing''}
\Else
    \State $\text{type} \gets$ \text{``constant''}
\EndIf
\For{$i = 1$ \textbf{ to } $l_S - 1$}
    \State $t_i \gets t_{\text{initial}} + i - 1$ \Comment{In 1-based indexing languages}
    \State $m \gets S_t[t_i+1] - S_t[t_i]$
    \If{$\text{type} = $ \text{``increasing''} \textbf{and} $m = 0$}
        \State $R \gets t_i$
        \State \text{append(increasing\_intervals,} $[L, R])$
        \State \text{append(intervals\_type, type)}
        \State $L \gets R$
        \State $\text{type} \gets$ \text{``constant''}
    \ElsIf{$\text{type} = $ \text{``constant''} \textbf{and} $m > 0$}
        \State $R \gets t_i$
        \State \text{append(constant\_intervals,} $[L, R])$
        \State \text{append(intervals\_type, type)}
        \State $L \gets R$
        \State $\text{type} \gets$ \text{``increasing''}
    \EndIf
\EndFor
\If{\text{(type} $ = $ \text{``increasing''} \textbf{and} $m > 0)$ \textbf{ or } \text{(type} $ = $ \text{``constant''} \textbf{and} $m > 0)$}
    \State $R \gets l_S$
    \State \text{append(increasing\_intervals,} $[L, R])$
    \State \text{append(intervals\_type, type)}
\Else
    \State $L \gets R$
    \State $R \gets l_S$
    \State \text{append(constant\_intervals,} $[L, R])$
    \State \text{append(intervals\_type, type)}
\EndIf

\State \Return \text{increasing\_intervals, constant\_intervals, intervals\_type}

\end{algorithmic}
\end{algorithm}





\begin{algorithm}[H]
\caption{LocalizedPhase}
\label{alg:localized_phase}
\begin{algorithmic}[1]
\Require{$l_{\text{deloc}}: \texttt{Float}$\text{, increasing\_intervals: }\texttt{Array(Array(Int))}}
\Ensure{$\text{localized\_phase: }\texttt{Array(Array(Int))}$}

\State \textbf{Initialization:}\text{interval1, interval2, current\_interval: }\texttt{Array(Int)}

\While{$\text{length(increasing\_intervals)} > 1$} \Comment{\text{increasing\_intervals} is treated as a stack data type}
    \State $\text{interval1} \gets \text{increasing\_intervals}[1]$
    \State $\text{interval2} \gets \text{increasing\_intervals}[2]$

    \If{$|\text{intervalo2}[1]-\text{intervalo1}[2]| \leq l_{\text{deloc}}$} \Comment{The intervals are combined and continue in the iteration process}
        \State $\text{current\_interval} \gets [\text{interva1}[1] \text{, interval2}[2]]$
        \State \text{pop(increasing\_intervals)}
        \State \text{pop(increasing\_intervals)}
        \State \text{pushfirst(increasing\_intervals, current\_interval)}
    \Else \Comment{The interval is assigned to the localized phase}
        \State $\text{current\_interval} \gets [\text{interva1}[1] + 1 \text{, interval1}[2]]$
        \State \text{pop(increasing\_intervals)}
        \State append(localized\_phase, current\_interval)
    \EndIf
\EndWhile

\If{$\text{increasing\_intervals} \textbf{ is NOT } \text{empty}$}
    \State $\text{current\_interval} \gets \text{increasing\_intervals}[1]$
    \State append(localized\_phase, current\_interval)
\EndIf

\State \Return \text{localized\_phase}

\end{algorithmic}
\end{algorithm}

\section{Relaxation of the perturbation amplitude $h(t)$ for $\beta=1$}\label{sec:relbeta1}

Let us slightly modify the ansatz (\ref{ansatz2}) as $h(t')=[\ln (t'+1)]^{\alpha}/(1+t')$ to avoid a divergence at $t'=0$.
We substitute this form for all $t'$ in the operator (\ref{opermem}),
\begin{eqnarray}
{\cal F}[h(t)]&=&\int_0^{t}dt'\frac{[\ln (t'+1)]^{\alpha}}{(1+t')(1+t-t')}\nonumber\\
&=&\frac{1}{t+2}\int_0^{t}dt'[\ln (t'+1)]^{\alpha}\left[\frac{1}{1+t'}+\frac{1}{1+t-t'}\right]\nonumber\\
&=&\frac{[\ln (t+1)]^{\alpha+1}}{(t+2)(\alpha+1)}+\frac{1}{t+2}\int_0^{t}dt'\frac{[\ln (t'+1)]^{\alpha}}{1+t-t'}\,.\label{a1}
\end{eqnarray}
We seek to evaluate the second integral in the rhs of Eq. (\ref{a1}). We fix once again a positive $\epsilon\ll1$ and a large time $t$ such that $\epsilon t\gg1$. Making the change $\tau=t-t'$,
\begin{eqnarray}
\int_0^{t}dt'\frac{[\ln (t'+1)]^{\alpha}}{1+t-t'}&=&\int_0^{\epsilon t}d\tau\frac{[\ln (1+t-\tau)]^{\alpha}}{1+\tau}
+\int_{\epsilon t}^t d\tau\frac{[\ln (1+t-\tau)]^{\alpha}}{1+\tau}\\
&=& A+B\,.
\end{eqnarray}
In the first contribution $A$, we expand the logarithm at first order in $\tau/t$ to obtain its leading behavior
\begin{equation}\label{A}
A\simeq (\ln t)^{\alpha} \int_0^{\epsilon t}\frac{d\tau}{1+\tau}\left(1-\frac{\alpha\tau}{t\ln t} \right)\simeq (\ln t)^{\alpha+1}+{\cal O}\left((\ln\epsilon) (\ln t)^{\alpha}\right)\,.
\end{equation}
As $[\ln(1+t-\tau)]^{\alpha}$ is a decreasing function of $\tau$, the second contribution $B$ fulfills
\begin{equation}
B< [\ln(1+t(1-\epsilon))]^{\alpha}\int_{\epsilon t}^{t}\frac{d\tau}{1+\tau}\to -(\ln \epsilon) (\ln t)^{\alpha}\ll (\ln t)^{\alpha+1}\,.
\end{equation}
Therefore $B\ll A$. Combining these results together into Eq. (\ref{a1}) yields the asymptotic behavior,
\begin{equation}\label{a6}
   {\cal F}[h(t)]\simeq \frac{(\ln t)^{\alpha+1}}{t} \left(\frac{1}{\alpha+1}+1\right)\,.
\end{equation}
Inserting Eq. (\ref{a6}) into Eq. (\ref{eqhbeta1}) gives the relation
\begin{equation}
    1=\left(1-\gamma P_0^{(st)}\right)\frac{\alpha+2}{\alpha +1}\,,
\end{equation}
which is equivalent to the result (\ref{alpha}) of the main text.


%

\end{document}